\documentclass[twocolumn]{aastex631}
\usepackage{float}
\usepackage{graphicx}
\usepackage{amsmath,amssymb}
\usepackage[section]{placeins}
\usepackage{hyperref}

\usepackage{soul}
\usepackage{threeparttablex}
\usepackage{multirow}

\begin{document}

\title[Are purely gaseous bars a myth?]
{Are purely gaseous bars in dwarf-irregulars a myth?}

\author{Anagha A G}
\affiliation{ Department of Physics, Indian Institute of Science Education and Research (IISER) Tirupati, Tirupati - 517507, India}

\author{Arunima Banerjee}
\affiliation{ Department of Physics, Indian Institute of Science Education and Research (IISER) Tirupati, Tirupati - 517507, India}

\begin{abstract}
About two-thirds of the galactic disks exhibit a central ellipsoidal stellar component called the bar, with or without a gaseous counterpart. However, there are a few dwarf galaxies with purely gaseous bars: NGC3741, NGC2915 and DDO168. This is a puzzle as gas is a collisional medium, and a gaseous bar is expected to be ripped off by shock waves. We study the formation of gaseous bars in these galaxies by constructing dynamical models constrained by stellar photometry and HI observations already available. We first analytically study the dynamical stability of the galactic disks against global $m=2$ perturbations. Our results indicate that the stellar and the gas disks are moderately unstable against these bar modes. Using N-body + hydrodynamical simulations employing $RAMSES$, we next find that a purely gaseous bar is formed in an oblate dark matter halo of vertical-to-planar axes ratio $c/a = 0.6 - 0.8$, with a relatively high-spin parameter $\Lambda = 0.04 - 0.07$, which survives for more than ten dynamical times. Further, the low values of our calculated Mach numbers $M=2-6$ of the gaseous medium comply with the survival of the gaseous bars, unaffected by shock waves. Interestingly, our simulations show the formation of a tiny stellar bar in each case. However, the temporal evolution of the change in angular momentum $L_z$ of the different disk components indicates the exchange of $L_z$ between the gas disk and the dark matter halo only; the $L_z$ of the stellar disk remained unchanged, indicating a weak stellar bar.

\end{abstract}


\keywords{galaxies: formation, galaxy: evolution,  galaxies: bars,  
galaxies: dynamics}



\section{Introduction}

Galactic disks may exhibit a central ellipsoidal stellar component known as the bar (\citeauthor{Eskridgeetal2000} \citeyear{Eskridgeetal2000},\citeauthor{MarinovaandJogee2007} \citeyear{MarinovaandJogee2007}, \citeauthor{Aguerrietal2009} \citeyear{Aguerrietal2009}, \citeauthor{Butaetal2015)} \citeyear{Butaetal2015)}). Stellar bars have been detected in about two-thirds of disk galaxies, including the Milky Way (\citeauthor{Aguerrietal2009} \citeyear{Aguerrietal2009}, \citeauthor{Butaetal2015)} \citeyear{Butaetal2015)}).   The quadrupole moment associated with the gravitational potential of the non-axisymmetric bar component helps in angular momentum transport (\citeauthor{LyndenBellKalnajs1972} \citeyear{LyndenBellKalnajs1972}, \citeauthor{LyndenBellKalnajs1972} \citeyear{LyndenBellKalnajs1972}), 
 and hence in gas inflow, thus feeding the galactic nuclei and fuelling star formation activity (\citeauthor{MartinetFriedli1997} \citeyear{MartinetFriedli1997};\citeauthor{Aguerri1999} \citeyear{Aguerri1999}, \citeauthor{SilvaLima2022} \citeyear{SilvaLima2022}, \citeauthor{Storchi_Bergmann_2019} \citeyear{Storchi_Bergmann_2019}, \citeauthor{kataria_2024} \citeyear{kataria_2024}, \citeauthor{garland2024} \citeyear{garland2024}). Besides, bars may be associated with spiral arms and may themselves drive the spiral density waves (\citeauthor{BinneyTremaine1987} \citeyear{BinneyTremaine1987}). In addition, bars may also lead to the heating of stellar disks (\citeauthor{Sahaetal2010} \citeyear{Sahaetal2010}) and thus play a fundamental role in the secular evolution of the galaxies.

Dynamical modeling of observed structure and kinematics indicates that the bar rotates about the galactic center with a constant angular speed called the pattern speed (\citeauthor{(Aguerrietal2015} \citeyear{(Aguerrietal2015}), with the stars therein streaming along highly eccentric orbits co-aligned with its major axis (\citeauthor{Weinberg1985} \citeyear{Weinberg1985}, \citeauthor{ContopoulosandGrosbol1989;}  \citeyear{ContopoulosandGrosbol1989;}, \citeauthor{Athanassoula1992;} \citeyear{Athanassoula1992;},\citeauthor{KormendyandKennicutt2004} \citeyear{KormendyandKennicutt2004}). The stellar dynamics in the galactic bar are non-linear, sustained by mostly chaotic orbits (\citeauthor{SellwoodWilkinson1993} \citeyear{SellwoodWilkinson1993})and, hence, it is challenging to study bar dynamics using analytical models. Bars are primarily modeled as density waves, supported by self-consistent stellar orbits (\citeauthor{BinneyTremaine1987} \citeyear{BinneyTremaine1987}). However, observational evidence may sometimes be at variance with theoretical predictions, which, in turn, challenge the foundations of current galaxy formation and evolution theories.

Most barred galaxies host a stellar bar, with or without a gaseous counterpart. However, there are a few gas-rich and dark matter-dominated dwarf galaxies that host purely gaseous bars, i.e., gaseous bars not accompanied by a stellar component. However, their occurrence is rare, and only three such galaxies have been reported in the literature so far: NGC 2915, NGC 3741 and DDO 168 (\citeauthor{Bureau1999}\citeyear{Bureau1999}; \citeauthor{Begum_2008}\citeyear{Begum_2008}; \citeauthor{PatraJog2019}\citeyear{PatraJog2019}). The occurrence of purely gaseous bars is a theoretical puzzle as, unlike stars, gas is a collisional medium. It hence cannot host self-consistent orbits sustaining the bar density wave (\citeauthor{ShuFrankH1992} \citeyear{ShuFrankH1992}). Numerical simulation studies have indicated that the supersonic gas motion, along with the shocks, occurs as a result of the response of the interstellar gas to the bar potential (\citeauthor{Sørensen1976} \citeyear{Sørensen1976}). If a non-axisymmetric instability is triggered in a gas component, shock waves may soon rip it off the gaseous bar structure. Therefore, it is a mystery how some dwarf irregular galaxies have sustained a purely gaseous bar. Besides, the physical mechanism stabilizing the stellar disk against stellar bar formation in such galaxies remains unknown. In NGC 2915, the origin of the gaseous bar, which is also accompanied by a spiral arm, was attributed to the quadrupolar potential of a slowly rotating triaxial dark matter halo. However, no dynamical analysis was done to establish the same (\citeauthor{Fathi_2009} \citeyear{Fathi_2009},  \citeauthor{stv1135} \citeyear{stv1135}), \citeauthor{Masset_2004} (\citeyear{Masset_2004}) carried out purely hydrodynamical simulations for NGC2915 to understand the bar and spiral structures observed in the gas disk. Interestingly, the spiral structure was found to be driven by a massive gas disk.

In this paper, we aim to address the puzzle of the existence of purely gaseous bars in the dwarf galaxies NGC3741, NGC2915, and DDO168. In each case, the bar is accompanied by spiral arms; however, those are not as promiscuous as compared to the bar. Therefore, we focus mainly on the bar dynamics in this paper. We construct dynamical models of these galaxies constrained by stellar photometry and HI interferometric observations available in the literature. We first analytically study the disk dynamical stability of these galaxies against global, non-axisymmetric $m=2$ perturbations using linear perturbation analysis (\citeauthor{global} \citeyear{global}). We then construct an N-body + hydrodynamical model using the publicly-available code DICE (Disk Initial Conditions Environment, Astrophysics Source Code Library, \citeauthor{Perret2016} \citeyear{Perret2016}) and evolve them with time RAMSES (\citeauthor{Tessier2002} \citeyear{Tessier2002}).

The rest of the paper is organized as follows: In \S 2, we describe our analytical models, and in \S 3, the N-body + hydrodynamical simulations. In \S 4, we present our sample; in \S 5, our input parameters; in \S 6, the results and discussion followed by conclusions in \S 7.

\section{Analytical Models: Disk Stability against global, non-axisymmetric perturbations} 

Bars can be considered as quasi-stationary density waves, and associated with long-lived $m=2$ global modes in the galactic disks. The reason behind the presence of purely gaseous bars in the three galaxies NGC3741, NGC2915, and DDO168 is unknown. To investigate this, we carry out the global mode analysis of their stellar as well as gas disks following \citeauthor{global} (\citeyear{global}). The disk is taken to respond to the gravitational potential of its self-gravity as well as the external force fields of the other disk components and the dark matter halo, as regulated by the Poisson equation. The halo and the other disk components are taken to be rigid.

The response of the one-component model is governed by the standard set of hydrodynamical equations, namely the momentum equations and the continuity equation, in addition to Poisson's equation. We present the same below for a 2-dimensional (razor-thin) disk in polar coordinates (r,$\phi$).

\begin{equation}
    \frac{\partial u}{ \partial t} + u \frac{\partial u }{ \partial r}
 + \frac{v }{ r} \frac{\partial u}{ \partial \phi}
 - \frac{v^2}{ r} =
 - \frac{1}{ \Sigma} \frac{\partial P_s }{ \partial r}
 - \frac{\partial }{\partial r} \Big( \Psi + \Psi _H \Big)
\
    \label{f1}
\end{equation}

\begin{equation}
    \frac{\partial v }{ \partial t}
 + u \frac{\partial v }{ \partial r}
 + \frac{v }{ r} \frac{\partial v }{ \partial \phi}
 + \frac{v u }{ r} =
 - \frac{1 }{ \Sigma r} \frac{\partial P_s }{ \partial \phi}
 - \frac{1 }{ r} \frac{\partial }{ \partial \phi} \Big( \Psi + \Psi_H \Big)
\ ,
    \label{f2}
\end{equation}

\begin{equation}
    \frac{\partial \Sigma }{ \partial t}
 + \frac{1 }{ r} \frac{\partial }{ \partial r} \Big( r\Sigma u \Big)
 + \frac{1 }{ r} \frac{\partial }{ \partial \phi} \Big( \Sigma v \Big) = 0
\ ,
    \label{f3}
\end{equation}

\begin{equation} 
    \Psi(r,\phi) = -
\int_{R_{in}}^{R_{out}} \int_{0}^{2 \pi}
{\Sigma(r^{\prime},\phi^{\prime}) r^{\prime} dr^{\prime} d \phi^{\prime}
 \over {\sqrt{r^{2}+r^{\prime 2} -
 2rr^{\prime} \cos (\phi - \phi^{\prime})}}}
\ ,
    \label{f4}
\end{equation}

Here, $u$ and $v$ are the radial and azimuthal velocities in the disk, $\Sigma$ the surface density, $P_s$ the pressure, $\Psi$ the gravitational potential due to the disk components, and $\Psi_H$ the same due to the dark matter halo. We perturb the equations (\ref{f1})-(\ref{f4}) by a global, non-axisymmetric dynamical instability of the form
\begin{equation}
    f(r)+f_1(r)e^{im\phi-i\omega t}
    \label{pert_global}
\end{equation}

\noindent with $f(r)$ denoting the physical quantity in the unperturbed case, $f_1(r)$ the amplitude of the corresponding perturbation, $m$ the azimuthal wave number and $\omega$ the complex frequency of the perturbation. Here, $f(r)$ can be either the surface density, the enthalpy $w$, or the gravitational potential. To remove the singularity at the origin $r = 0$, we introduce the
coordinate transformations of the perturbations in enthalpy, potential
, and surface density as \\
$$
w_1      = r^m \tilde{w}_1     ~, \quad
\Psi_1   = r^m \tilde{\Psi}_1  ~, \quad
\sigma_1 = r^m \tilde{\sigma}_1~.
$$\\

\noindent The perturbation in enthalpy $\omega_1$ is related to the perturbation in surface density $\Sigma_1$ as $\omega_1 = \frac{C_r^2}{\Sigma}\Sigma_1$, where $C_r$ is the radial velocity dispersion. $C_r$ is determined from the vertical velocity dispersion $C_z$ by assuming $C_z/C_r = 0.5$ (\citeauthor{Binney2008} \citeyear{Binney2008}). Here, the surface density $\Sigma$ is given by
\begin{equation}
\Sigma    = \Sigma_0\mathrm{exp}(-r/h_\sigma)((1- r/R_{OUT})^2)^5 
\end{equation}
where $\Sigma_0$ is the central surface density, $h_\sigma$ the radial disk scale length, and $\mathrm{R_{OUT}}$ is the truncation radius of the stellar (gas) disk. The vertical velocity dispersion $C_z$ is modeled as
\begin{equation}
C_z  = C_{z}(0)\mathrm{exp}(-r/(2h_\sigma))((1- r/R_{OUT})^2)^{2.5}
\end{equation}
Here $\mathrm{C_{z}(0)}$ the central vertical dispersion. 
After linearizing equations (\ref{f1})--(\ref{f4}) with respect to the perturbed
quantities, and eliminating perturbed velocities, one obtains an equation of the form

\begin{equation}
    \frac{d^2(\Tilde{\omega_{1}} + \Tilde{\psi_{1}})}{dr^2}+A\frac{d(\Tilde{\omega_{1}} + \Tilde{\psi_{1}})}{dr}+B(\Tilde{\omega_{1}} + \Tilde{\psi_{1}})-\frac{D}{C_{r}^2}\Tilde{\omega_{1}}=0
    \label{eglobal_mode}
\end{equation}
where 

\begin{equation}
    A=\frac{2m+1}{r}+\frac{1}{\Sigma}\frac{d\Sigma}{dr}-\frac{1}{D}\frac{dD}{dr}
\end{equation}

\begin{equation}
    B=\frac{m}{r}((\frac{1}{\Sigma} \frac{d\Sigma}{dr})(1-\frac{2\Omega}{\omega-m\Omega})-\frac{2}{\omega-m\Omega}\frac{d\Omega}{dr})
\end{equation}
\begin{equation}
    D=\kappa^2-(\omega-m\Omega)^2
\end{equation}

\noindent Here $\kappa$ is the epicyclic frequency given by
\begin{equation}
    \kappa^2 = \frac{1}{r^3} \frac{d}{dr} (r^4 \Omega^2)
    \label{kappa}
\end{equation}
$\Omega$ is the angular velocity of the disk. The perturbed gravitational potential ($\tilde{\Psi}_1$) is given by

\begin{equation}
\begin{split}
    \tilde{\psi_1}(r) = -2 \pi \frac{\Gamma(m + 1/2)}{\Gamma(1/2) \Gamma(m+1)} \\
    \bigg(\int_0^r \Sigma_1(r')\frac{r'}{r}^{2 m+1}  
        F \left(0.5, m+0.5, m+1, \frac{r'}{r}^2 \right)dr'+ \\
    \int_r^{R_{out}} \Sigma_1(r') F\left(0.5, m+0.5, m+1, \frac{r'}{r}^2\right)dr'\bigg) 
\\
\end{split}
\label{perturbed_gravitational_potential}
\end{equation}

\noindent where $F$ is a hypergeometric function. The inner boundary condition follows from the regularity of solutions at the center of the disk:
\begin{equation}
    \begin{split}
        \frac{d}{dr} (\tilde{w}_1 + \tilde{\Psi}_1)
    + \frac{m}{2 m + 1}
    ( ( \frac{1}{\Sigma} \frac{d\Sigma}{dr}
    - \frac{1}{D}\frac{dD}{dr} ) \\
        \times ( 1 - \frac{2\Omega}{\omega - m \Omega} )
    -\frac{2}{\omega - m \Omega}\frac{d\Omega}{dr})
        (\tilde{w}_1 + \tilde{\Psi}_1)\\
    = 0 
    \end{split}
    \label{bound_1}
\end{equation}
Likewise, at the outer boundary where $C_r^2 = 0$, we require

\begin {equation}
    \frac{1}{\sigma}\frac{dP_s}{dr}
\left( \frac{d}{dr} (\tilde{w}_1 + \tilde{\Psi}_1)
+ \frac{m}{r} (\tilde{w}_1 + \tilde{\Psi}_1) \right)
- D \tilde{w}_1 = 0 
    \label{bound_2}
\end{equation}

The eigenvalue problem, defined by equations (\ref{eglobal_mode})–(\ref{perturbed_gravitational_potential}) and boundary conditions (\ref{bound_1}) and (\ref{bound_2}), is solved numerically, which yields an eigenfrequencies $\omega$ of the $m=2$ global mode. The growth rate and the pattern speed of the mode are $2\pi/\mathrm{Im}(\omega$) and Re($\omega$)/m, respectively. We analyze the response of stellar and the gas disks to the $m=2$ mode perturbation.

\section{N-body + hydrodynamical simulations}
\subsection{Generating Initial Conditions}
Our dynamical model consists of a gravitationally coupled stellar disk, a gas disk, and a triaxial dark matter halo. Using the publicly-available code DICE (Disk Initial Conditions Environment, Astrophysics Source Code Library, \citeauthor{Perret2016} \citeyear{Perret2016}), we set up the initial conditions for the galaxy. 

The surface brightness profiles of the galactic disks exhibit exponential distributions over a large dynamic range in radii (\citeauthor{deVaucouleurs1959} \citeyear{deVaucouleurs1959}, \citeauthor{Freeman1970} \citeyear{Freeman1970}). In our simulation, we use an exponential density profile to model both our stellar disk and gas disk, as given by

 \begin{equation}
     \rho_g(r,z)={\frac{M_g}{2 \pi R_d^2}}\mathrm{exp}\Bigg({-\frac{r}{R_d}}\Bigg){\mathrm{exp}}\Bigg({-\frac{z}{h_d}\Bigg)} 
     \label{disk}
 \end{equation}

\noindent where $R_d$ and $h_d$ are the radial disk scale length and vertical scale height of the disk component, respectively. For the gas disk, we choose a disk scale length large enough such that the surface density at all radii is roughly equal to the average of the observed gas surface density. The velocity dispersion of the gas disk  ${{\sigma}}$ is regulated by the Toomre
$Q$ parameter (\citeauthor{Toomre} \citeyear{Toomre}) given by 
\begin{equation}
    Q = \frac{\sigma \kappa}{\pi G \Sigma} 
\end{equation}
\noindent  Q $>$ 1 gives a stable disk. $\Sigma$ is the gas surface density obtained by integrating the mass density along the $z$ direction. A dark matter halo characterized by a triaxial potential given  by

\begin{equation}
    \begin{split}
        & \Phi_{\mathrm{halo}}(x,y,z)= \\
        & 2\pi abc \rho_0 r_s^2 \int_{0}^{\infty} \frac{s(\tau)}{r_s+s(\tau)} \frac{d\tau}{\sqrt{(a^2+\tau)(b^2+\tau)(c^2+\tau)}}
        \label{eswing_ampli}
    \end{split}
\end{equation}

\noindent where $a$, $b$, $c$ are the axis ratios and 
\begin{equation}
    s(\tau)=\frac{x^2}{a^2+\tau}+\frac{y^2}{b^2+\tau}+\frac{z^2}{c^2+\tau}
\end{equation}
 is the coordinate transformation equation from $r\rightarrow s$. 
 We get an oblate halo for a = 1, b = c $<$ 1. We get a prolate halo for a = 1, b = c $>$ 1. For $a=b=c$, this potential reduces to the NFW potential (\citeauthor{NFW} \citeyear{NFW})
 \begin{equation}
     \rho_{halo}=\frac{\rho_s}{x(1+x)^2}
 \end{equation}

\noindent where $x=r/r_s$, $\rho_s$ and $r_s$ are free parameters. The input parameters can be varied to regulate the effect of the halo on the disk dynamics. In particular, the ratio of the principal axes will govern the degree of triaxiality of the halo (\citeauthor{Peñarrubiaetal2009} \citeyear{Peñarrubiaetal2009}).

The rotation of the halo is governed by the spin parameter (\citeauthor{peebles1969} \citeyear{peebles1969}) given below:
\begin{equation}
    \lambda=\frac{J\sqrt{E}}{GM^{5/2}}
\end{equation}

\noindent where $J$ is angular momentum within virial radius, $E$ is the total energy and $M$ is the virial mass. 

\subsection{Running the simulation} 
Using the publicly available N-body + hydrodynamical code RAMSES (\citeauthor{Tessier2002} \citeyear{Tessier2002}), we then evolve the gravitationally-coupled distribution of the stellar, gas, and dark matter particles with time for over ten dynamical times. The dynamical time is the time taken by a particle at the periphery of the galactic disk to complete one revolution about the galactic center, and a galaxy is supposed to attain a state of equilibrium in a few dynamical times. In a galaxy, stars and dark matter constitute a collision-less medium, whereas gas includes a collisional and dissipative medium. Therefore, the stellar and the dark matter components are modeled using N-body dynamics while gas is modeled as a hydro-dynamical component (\citeauthor{Bodenheimeretal2007} \citeyear{Bodenheimeretal2007}). While running the simulation, we save the snapshots, i.e., both components' positions and velocities, at a time interval.

\subsection{Methods of Analysis} 

Using the publicly-available "pynbody" code (\citeauthor{Pontzen2013} \citeyear{Pontzen2013}), we generate (i) the rotation curve and (ii) the gas radial surface density profiles to check if they match their observed counterparts.

We identify the gaseous bars in our simulated galaxy using the \emph{Fast Fourier Transform} method, which represents the amplitudes of different Fourier components in the surface brightness distribution of the galaxy (\citeauthor{(OhtaHamabeandWakamatsu1990} 
\citeyear{(OhtaHamabeandWakamatsu1990}, \citeauthor{PetersandKuziodeNaray2019}, \citeyear{PetersandKuziodeNaray2019}). The surface brightness of a gas disk can be written as

\begin{equation}
    \Sigma(r,\theta)=\sum_{m=0,1,2,3...} A_{m}(r)e^{im\theta}
\end{equation}
where $A_{m}$ is the amplitude of the ${m}^{th}$ Fourier mode, m is the mode number and $\theta$ is the azimuthal angle. The amplitude of the $m=2$ mode in the Fourier amplitude plot indicates the strength and prominence of the bar. \\

\textbf{Bar-Strength:} The bar-strength is given by 
\begin{equation}
    I_{b}/I_{ib}=\frac{A_0+A_2+A_4+A_6}{A_0-A_2+A_4-A_6}
\end{equation}

\noindent where $A_m$ is the amplitude of $m^{th}$ Fourier component from the Fast Fourier Transform (FFT) of a snapshot (\citeauthor{(OhtaHamabeandWakamatsu1990} 
\citeyear{(OhtaHamabeandWakamatsu1990}), $I_b$ and $I_{ib}$ corresponding to the bar and inter-bar intensity respectively. The above expression can also be used to determine the strength of the spiral arm. \\

\textbf{Pattern Speed}
The pattern speed of a bar refers to the angular velocity of the bar (or a spiral arm), which rotates as a rigid body in general. There are several methods in the literature to estimate the pattern speed. To determine the pattern speed of a bar (or a spiral arm) from the Fourier amplitude, the primary approach involves measuring the phase shift of the $m=2$ component with respect to the azimuthal angle of the galaxy. The phase angle at a radius $r$ is then given by
\begin{equation}
    \Phi_{m}=\mathrm{tan}^{-1}\frac{Im(A(r))}{Re(A(r))}
\end{equation}
The pattern frequency is then determined from the phase angle using the following equation. 

\begin{equation}
    \Omega_{pm}=\frac{1}{m}\frac{d \Phi_{m}}{dt}
    \label{epattern_freq}
\end{equation}

{\textbf{Length \& eccentricity of the bar}}
The geometry of the bar, including its position angle, ellipticity, and length, are determined from the isophotal ellipse fits (\citeauthor{Bradleyetal2020}\citeyear{Bradleyetal2020}) obtained using the photutils.isophote module of the astropy package (\citeauthor{larry_bradley_2022_6825092}\citeyear{larry_bradley_2022_6825092}). We generate radial profiles of ellipticity (e) and position angle (PA) from the ellipse fits, which vary with the galactocentric radius. The length of the bar is indicated by the semi-major axis length (sma), where the ellipticity reaches its maximum value (\citeauthor{WP} \citeyear{WP}, \citeauthor{Wozniaketal1995} \citeyear{Wozniaketal1995}, \citeauthor{jung_1997} \citeyear{jung_1997}, \citeauthor{Butaetal1998} \citeyear{Butaetal1998}, \citeauthor{laine_2002} \citeyear{laine_2002}, \citeauthor{Sheth_2003} \citeyear{Sheth_2003}, \citeauthor{Menéndez-Delmestre_2007} \citeyear{Menéndez-Delmestre_2007}, \citeauthor{MarinovaandJogee2007} \citeyear{MarinovaandJogee2007},   \citeauthor{Guoetal2023} \citeyear{Guoetal2023}). In our work, we employ this approach to measure the bar length.
\begin{table}
	\centering
	\caption{Physical properties of the galaxy sample}
	\label{table:1}
	\begin{tabular}{lccc} 
		\hline
		 Properties & NGC3741 & NGC2915 & DDO168 \\ 

		\hline
		$V_{\mathrm{sys}} (kms^{-1})$ $^\mathrm{1}$ & $229.1$ & $468\pm5$ & $192.6\pm1.2$ \\
		  D (Mpc) $^\mathrm{2}$& 3.24 &  $5.3\pm1.6$ & 4.3 \\
		$M_{*} (10^{8}M_{\odot})$ $^\mathrm{3}$ & $0.14$ &  $4.8$ & $0.59$ \\
            $R_{\mathrm{d*}} (kpc)$ $^\mathrm{4}$ &0.3&0.6 & 0.6\\
            SFR ($10^{-3}M_\odot yr^{-1}$)&3.4& 50 & 5.3 \\
            $M_{\mathrm{H1}}(10^{8}M_{\odot})$ $^\mathrm{5}$ & $1.69$ &  $9.58$ &   $2.6$\\
            $R_{\mathrm{dg}} (kpc)$ $^\mathrm{6}$ &2 &5.4 & 0.8\\
            ${\mbox{H\,{\sc i}}}$ Bar length&1.2&1&1.8\\

		\hline
	\end{tabular}
    \begin{tablenotes}
    \raggedright
       \item $^\mathrm{1}$ Systematic velocities of the three galaxy samples\\
       $^\mathrm{2}$ Distance of the galaxy from the line of sight in Mpc\\
       $^\mathrm{3}$ Total mass  of the stellar disk\\
       $^\mathrm{4}$ Scale radius of the stellar disk\\
       $^\mathrm{4}$ SFR is the star formation rate\\
       $^\mathrm{5}$ Total mass of the gas disk\\
       $^\mathrm{6}$ Scale radius of the gas disk\\

    \end{tablenotes}
\end{table}

\begin{table}
    \centering
    \caption{Input parameters: Global mode study}
    \begin{tabular}{lccc}
    \hline
     & NGC3741& NGC2915& DDO168 \\
    Gas disk  \\
    \hline
        $\Sigma_0$ $^\mathrm{1}$  & $\mathrm{16 \ M_\odot \ pc^{-2} }$ & $\mathrm{30 \ M_\odot \ pc^{-2} }$& $\mathrm{20 \ M_\odot \ pc^{-2} }$ \\
        $R_d$ $^\mathrm{2}$& $\mathrm{2 \ kpc}$ & $\mathrm{5.4 \ kpc} $& $\mathrm{0.8 \ kpc} $\\
        $\sigma_{HI}$ $^\mathrm{7}$ & $10 \ \mathrm{kms^{-1}}$& $10 \ \mathrm{kms^{-1}}$& $15 \ \mathrm{kms^{-1}}$ \\
	\hline
        Stellar disk & \\
        \hline	
        $\Sigma_0$ $^\mathrm{3}$  & $\mathrm{12.2 \ M_\odot \ pc^{-2} }$ & $\mathrm{17 \ M_\odot \ pc^{-2} }$& $\mathrm{13 \ M_\odot \ pc^{-2} }$ \\
        $R_d$ $^\mathrm{4}$ & $\mathrm{0.3 \ kpc}$ & $\mathrm{0.6 \ kpc} $& $\mathrm{0.6 \ kpc} $\\
        $h_z$ $^\mathrm{5}$ & $\mathrm{R_d/6}$ & $\mathrm{R_d/6}$& $\mathrm{R_d/6}$\\
        $\sigma_z(0)$ $^\mathrm{6}$   & $5 \ \mathrm{kms^{-1}}$& $5 \ \mathrm{kms^{-1}}$& $5 \ \mathrm{kms^{-1}}$ \\
	\hline
        Halo& \\ 
        \hline
        $\rho_0$& 0.0248 $\mathrm{M_\odot  pc^{-3}}$ & 0.1 $\mathrm{M_\odot  pc^{-3}}$ & 0.01 $\mathrm{M_\odot  pc^{-3}}$ \\
        $R_c$  & 2.97 kpc & 1.3 $\pm$ 0.2 kpc & 2.81 kpc \\
 \hline

        \hline
    \end{tabular}
    \begin{tablenotes}
    \raggedright
       \item $^\mathrm{1}$ Central ${\mbox{H\,{\sc i}}}$ surface density\\
       $^\mathrm{2}$ Exponential ${\mbox{H\,{\sc i}}}$ disk scale length\\ 
          $^\mathrm{3}$ Central stellar surface density in B-band\\
          $^\mathrm{4}$ Exponential stellar disk scale length\\
          $^\mathrm{5}$ Stellar scale height \\
          $^\mathrm{6}$ Central stellar velocity dispersion\\
          $^\mathrm{7}$ Stellar vertical velocity dispersion for scale height $\sim \ R_d/6$

    \end{tablenotes}
\end{table}

\begin{table}
	\centering
	\caption{Input parameters for the N-body + hydrodynamical simulations of our sample galaxies}
	\label{input_parameters}
	\begin{tabular}{lccc} 
		\hline
		  & NGC3741 & NGC2915 & DDO168\\
        \hline
        Dark matter halo& \\ 
        \hline
        $ \mathrm{V_{200}}(km/s)^\mathrm{1}$&  35.3 & 66.4 & 34.4 \\
		$\Lambda ^\mathrm{2}$& 0.07&0.04&0.07 \\
		$c/a ^\mathrm{3}$ & 0.8&0.8&0.6 \\
		$C $ $^\mathrm{4}$ &$11.2\pm0.8$&12.9 & 16.8\\
        \hline
        Gas disk& \\ 
        \hline
        $m\%_{g} ^\mathrm{5}$& 3.9&5.14 &7.63 \\
        $h_zg (kpc) ^\mathrm{6}$ &0.33 & 1.8&0.13 \\
        \hline
        Stellar disk& \\ 
        \hline
        $m\%_{*} ^\mathrm{7}$& 0.3&0.27 & 1.64\\
        $h_z* (kpc) ^\mathrm{8}$ &0.06 & 0.1&0.1 \\
        \hline

		\hline
	\end{tabular}
    \begin{tablenotes}
    \raggedright
       \item  $^\mathrm{1}$ virial velocity\\
       $^\mathrm{2}$ spin value of dark matter halo\\
         $^\mathrm{3}$  axis ratio of dark matter halo\\
       $^\mathrm{4}$ Concentration parameter of the NFW dark matter halo\\
       $^\mathrm{5}$ gas mass fraction.\\
       $^\mathrm{6}$ Scale height of  gas disk \\
       $^\mathrm{7}$ stellar mass fraction\\
       $ ^\mathrm{8}$ scale height of stellar disk \\
    \end{tablenotes}
\end{table}

\begin{figure*}[h]
\begin{center}
    \includegraphics[scale=0.55]{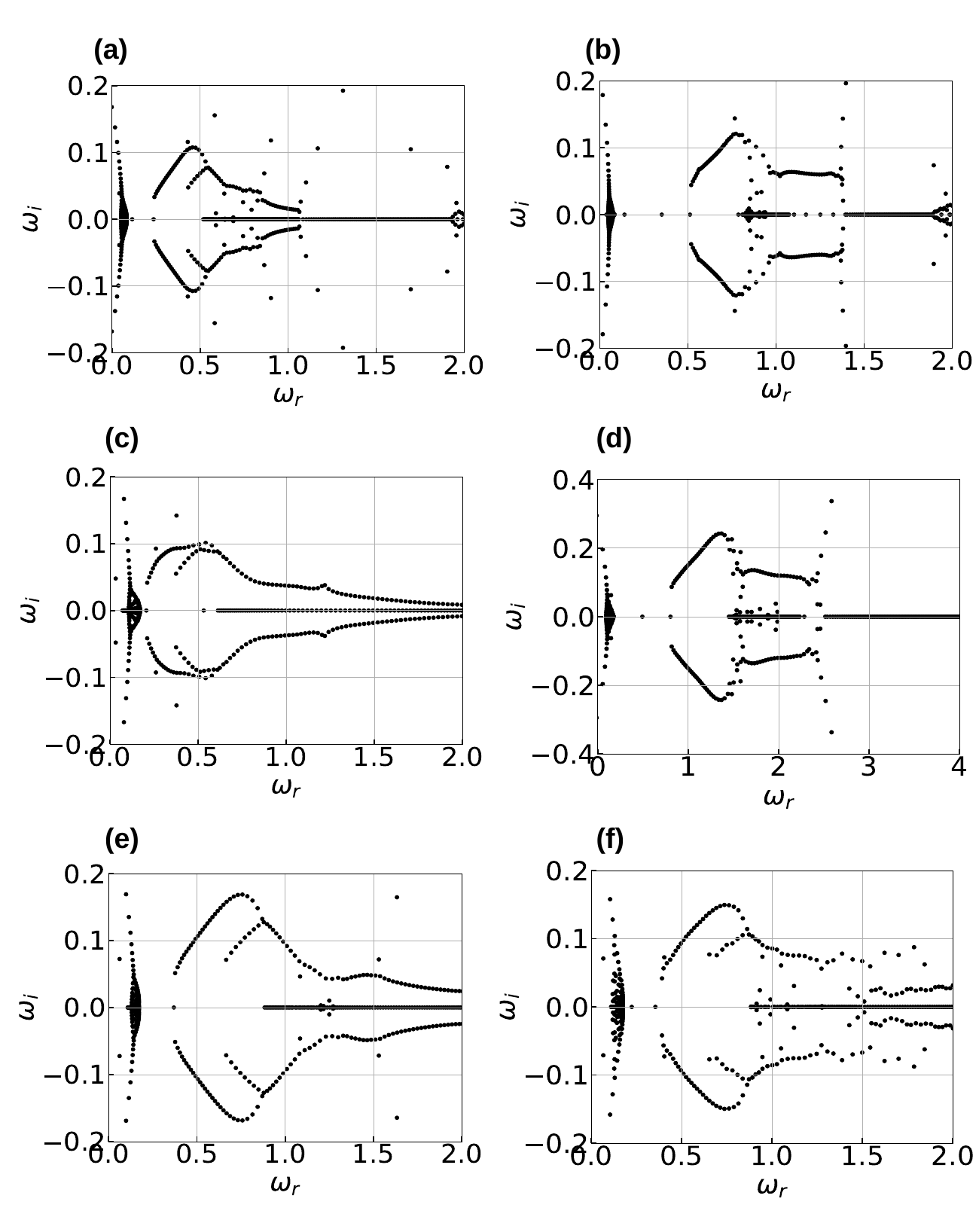}
\end{center}
\caption{The eigenspectrum plot of dominant m=2 global modes from the global mode analysis.
Top Row: NGC3741 Middle Row: NGC2915 Bottom Row: DDO168. The left and right panels correspond to the gas and the stellar disks, respectively.}
\label{global_mode}
\end{figure*}

\begin{figure*}[h]
    \centering
    \includegraphics[scale=0.3]{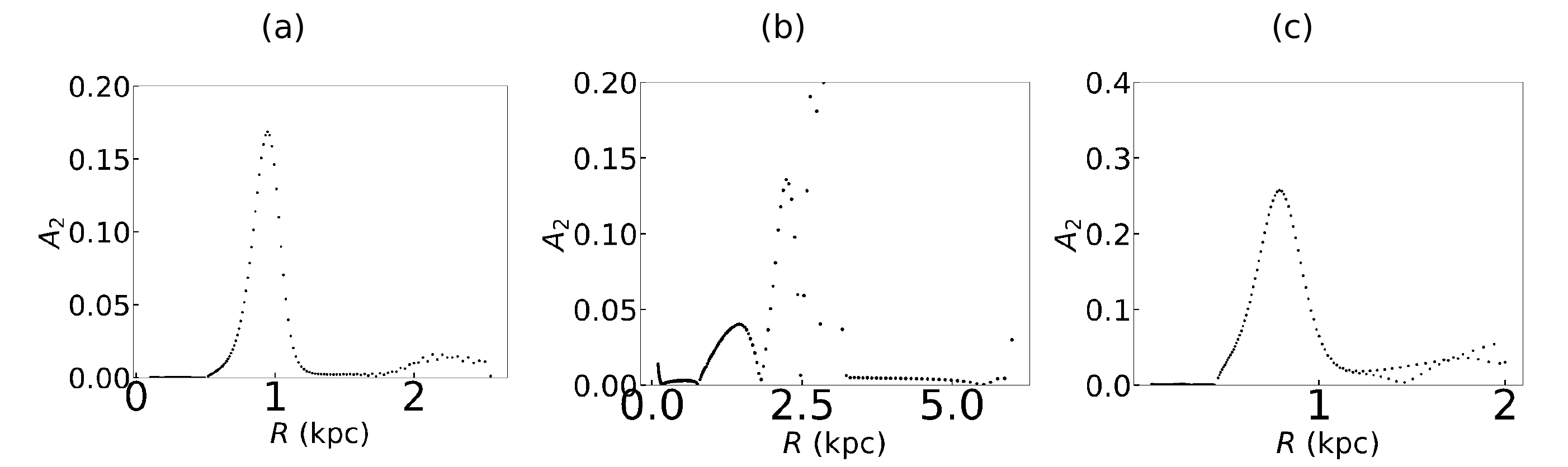}
    \caption{Amplitude of global m=2 mode perturbation as a function of galacto-centric radius at 1 Gyr in the gas disk of a) NGC3741, b)NGC2915, and c) DDO168, as obtained from global mode analysis.}
    \label{globalp}
\end{figure*}

\begin{figure*}
\begin{center}
    \includegraphics[scale=0.5]{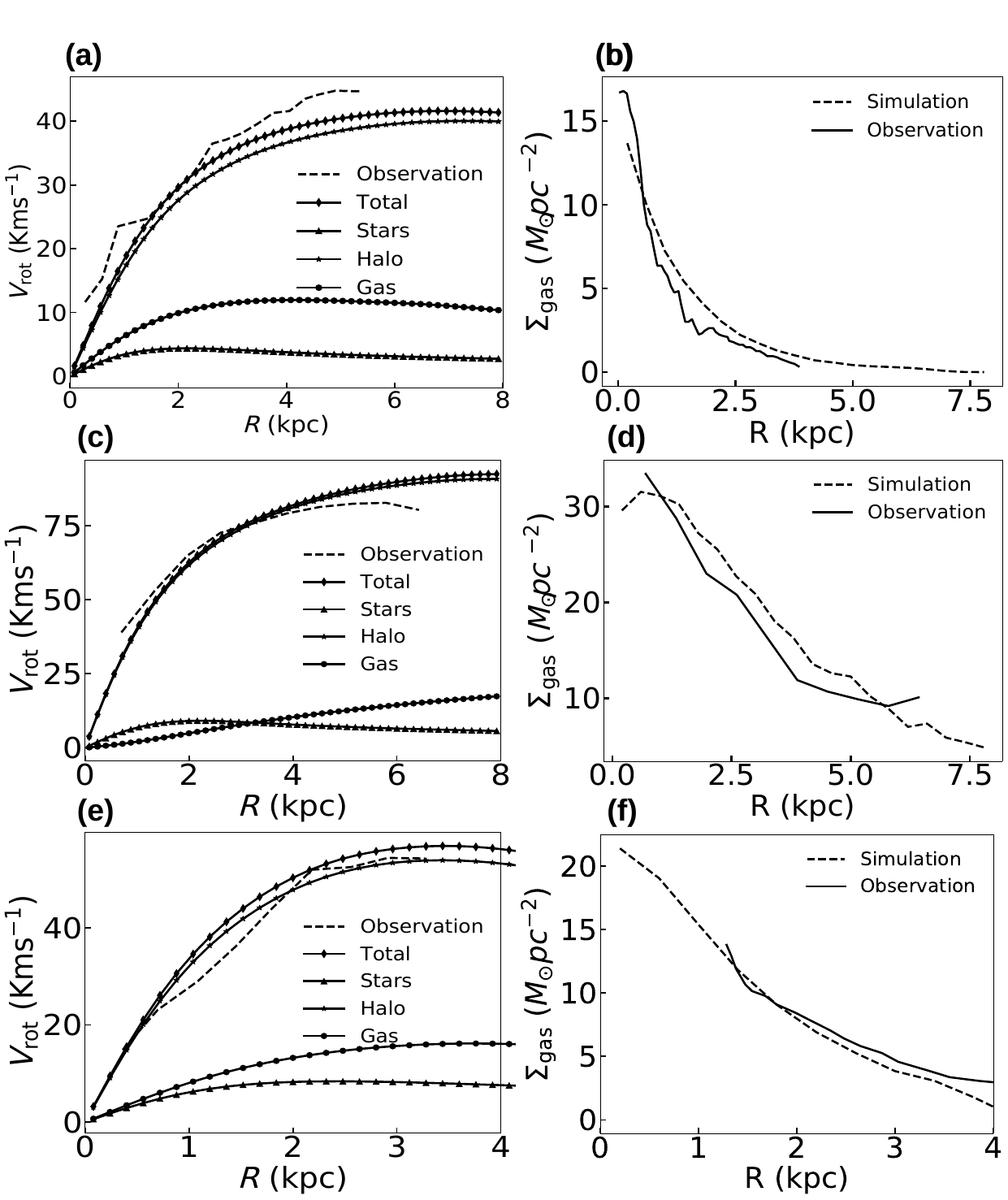}
\end{center}
\caption{The observed rotation curve (Left) and the surface density profiles (Right) as compared with the initial conditions of simulated snapshots. Top Row: NGC3741 Middle Row: NGC2915 Bottom Row: DDO168}
\label{physical_paramater}
\end{figure*}

\begin{figure*}[h]
\begin{center}
    \includegraphics[scale=0.2]{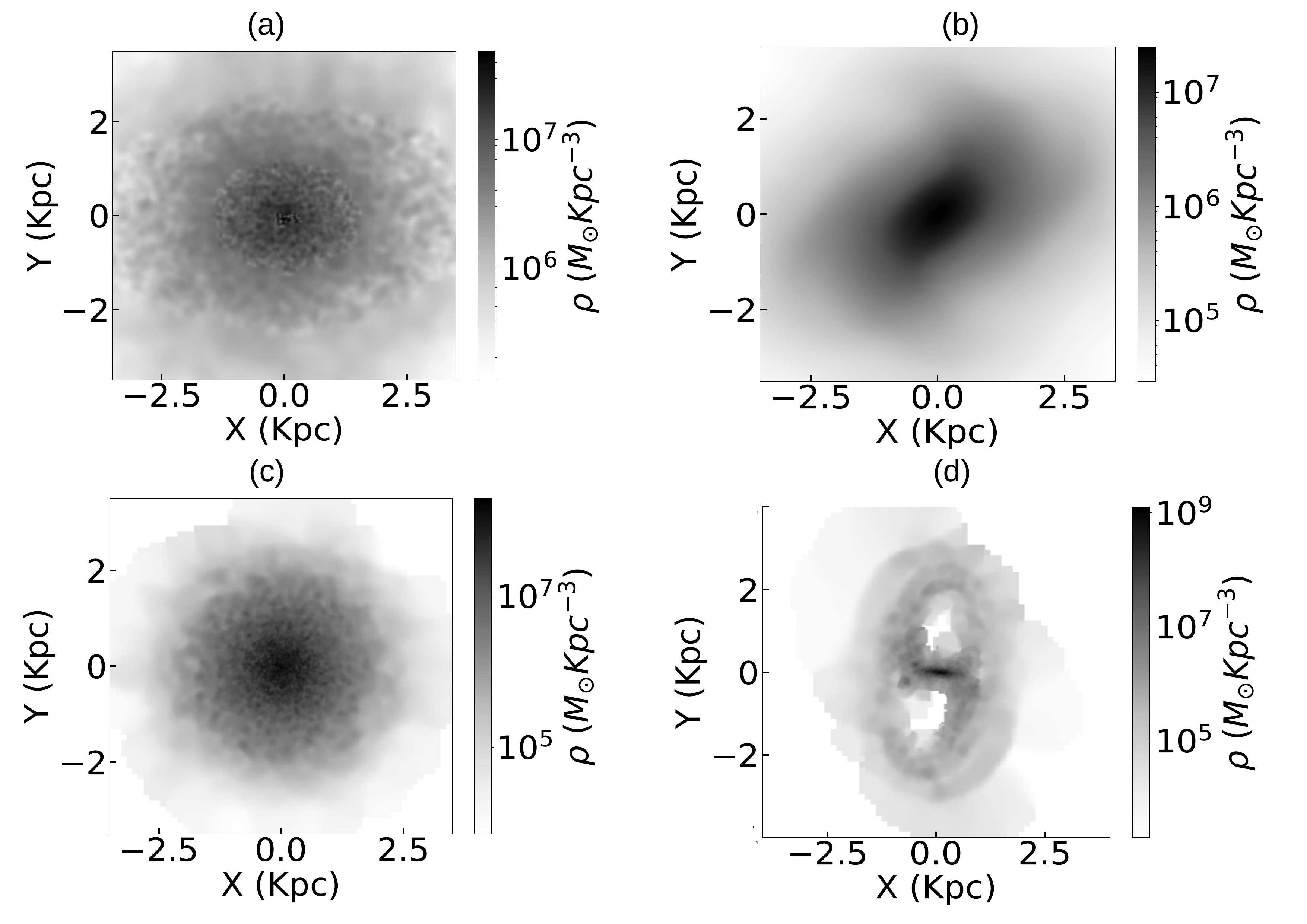}
\end{center}
\caption{Disk snapshots from RAMSES for the gas and stellar disks of NGC3741 at different epochs. (Top panel) the gas disk: (a) initial snapshot (time = 0 GYr) (b) snapshot at a later time 1.13 GYr (9.8 * dynamical time of galaxy) when the bar has appeared in the gas disk and (Bottom Panel) the stellar disk, both in face-on orientations : (c) initial snapshot (time = 0 GYr) (d) snapshot at a later time 1.13 GYr (9.8 * dynamical time of galaxy) when the bar has appeared in the gas disk.}
\label{disk_snapshots_NGC3741}
\end{figure*}

\begin{figure*}[h]
\begin{center}
    \includegraphics[scale=0.2]{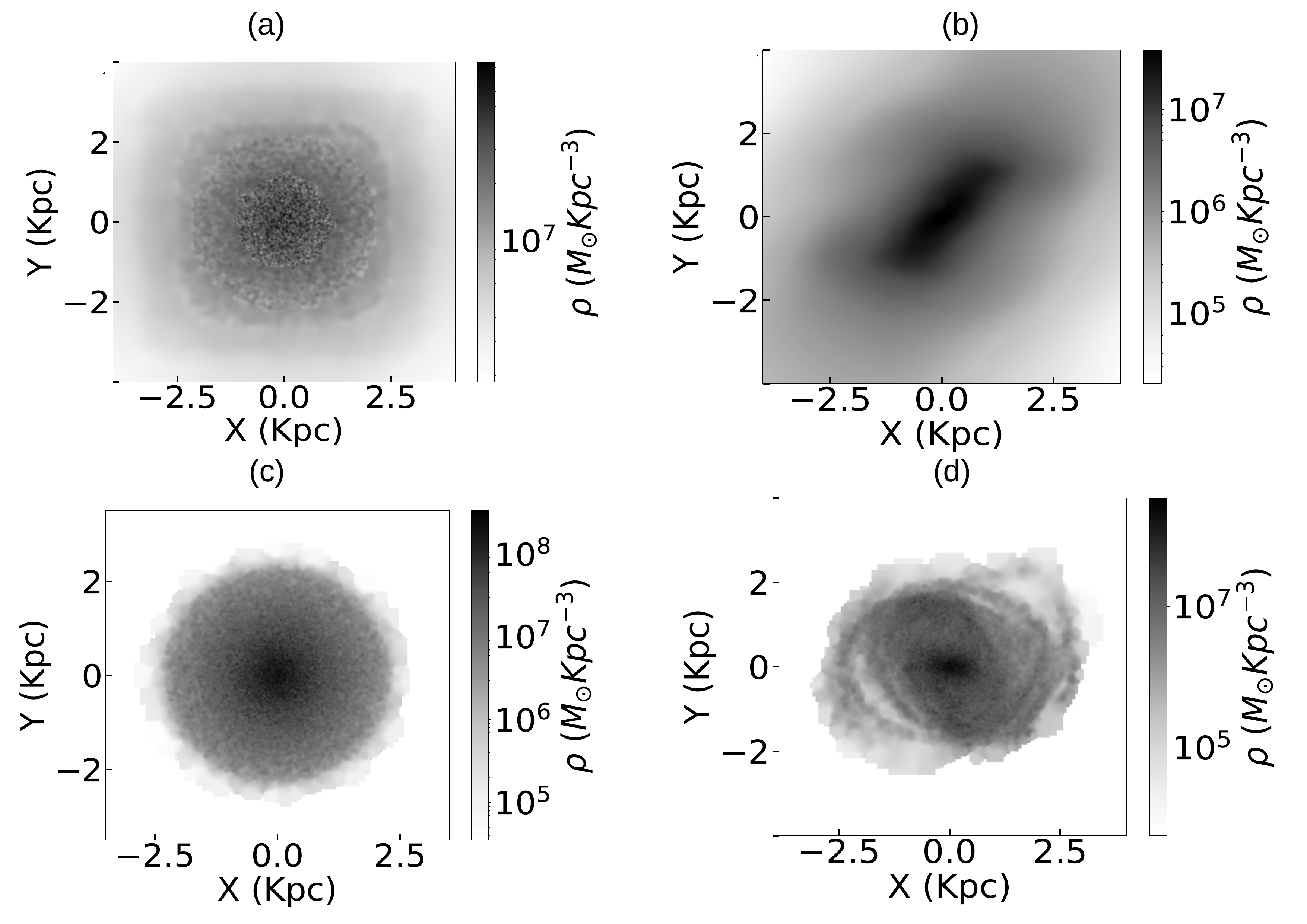}
\end{center}
\caption{Disk snapshots from RAMSES for the gas and stellar disks of NGC2915 at different epochs in face-on orientations. (Top Panel) The gas disk : (a) initial snapshot (time = 0 GYr) (b) snapshot at a later time 0.29 GYr (3 * dynamical time of galaxy) when the bar has appeared in the gas disk and (Bottom Panel) the stellar : (c) initial snapshot (time = 0 GYr) (d) snapshot at a later time 0.29 GYr (3 * dynamical time of galaxy) when the bar has appeared in the gas disk.}
\label{disk_snapshots_NGC2915}
\end{figure*}

\begin{figure*}[h]
\begin{center}
    \includegraphics[scale=0.2]{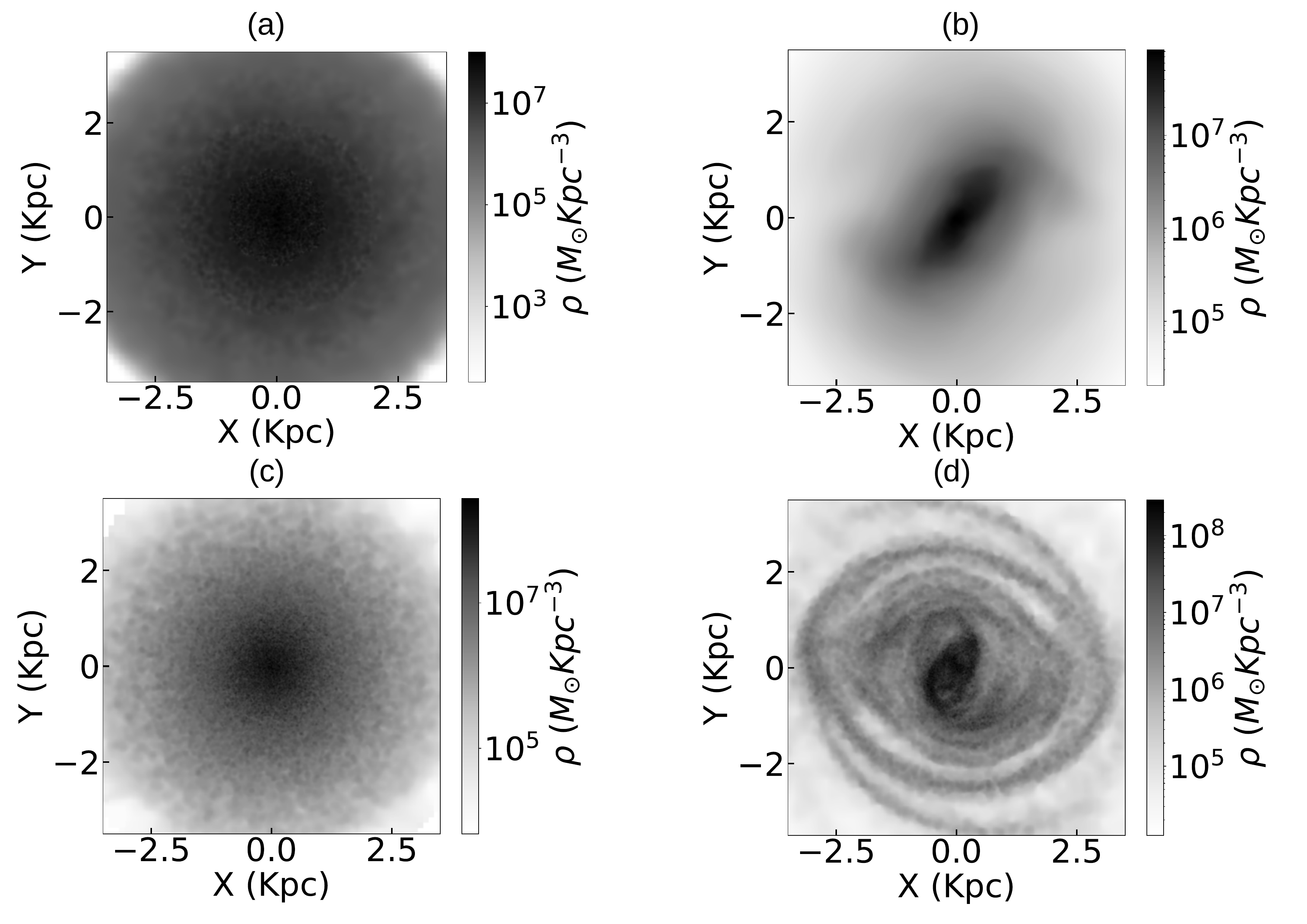}
\end{center}
\caption{Disk snapshots from RAMSES for the gas and stellar disks of DDO168 at different epochs in face-on orientations. (Top Panel) The gas disk : (a) initial snapshot (time = 0 GYr) (b) snapshot at a later time 0.6 GYr (6 * dynamical time of galaxy) when the bar has appeared in the gas disk and (Bottom Panel) the stellar disk :  (c) initial snapshot (time = 0 GYr) (d) snapshot at a later time 0.6 GYr (6 * dynamical time of galaxy) when the bar has appeared in the gas disk.}
\label{disk_snapshots_DDO168}
\end{figure*}

\section{Our galaxy sample} \label{samples}
Only three galaxies have been reported to exhibit purely gaseous bars and spiral arms: NGC2915 (\citeauthor{Bureau1999} \citeyear{Bureau1999}), NGC3741 (\citeauthor{Begum_2008} \citeyear{Begum_2008}, \citeauthor{Banerjee2013} \citeyear{Banerjee2013}), and DDO168 (\citeauthor{PatraJog2019} \citeyear{PatraJog2019}). NGC2915 is a blue compact dwarf galaxy with an extended ${\mbox{H\,{\sc i}}}$ disk. The gas disk features a short central bar and spiral arms, extending the optical component. Using the Tremaine-Weinberg method, \citeauthor{Bureau1999} (\citeyear{Bureau1999}) determined the pattern speed of the ${\mbox{H\,{\sc i}}}$ bar and the spiral arms of NGC2915 to be of $\Omega_p=0.21\pm0.06$km s$^{-1}$ arcsec$^{-1}$ ($8.0\pm2.4$~km~s$^{-1}$~kpc$^{-1}$ for $D=5.3$~Mpc). NGC3741 is a gas-rich ($M_{\mbox{H\,{\sc i}}}/L_B = 6.26 $) dwarf irregular galaxy with the 
${\mbox{H\,{\sc i}}}$ disk extending to about 8.3 times its Holmberg radius and marked by the presence of purely gaseous substructures like a bar and a spiral arm. \citeauthor{Banerjee2013} (\citeyear{Banerjee2013}) found the pattern speed of the \ion{H}{1} bar in NGC 3741 to be $\Omega_p$ = 17.1 ± 3.4 km s$^{-1}$ kpc$^{-1}$. Further, the ratio of the corotation radius to the bar semi-major axis was found to be 1.6 $\pm$ 0.3, indicating a slow bar. Recently, \citeauthor{PatraJog2019} (\citeyear{PatraJog2019}) studied the total intensity maps of the VLA LITTLE-THINGS galaxies and identified a weak \ion{H}{1} bar of strength 0.2 and radius one kpc in the nearby dwarf irregular galaxy DDO168. The dimensionless ratio, $R_L/R_b$, is $>$ 2.1 ( $R_L$ is the corotation radius, and $R_b$ is the bar length), indicating a slow bar in DDO 168.  

\section {Input Parameters}

NGC3741, NGC2915, and DDO168 are located at a distance of approximately  3.24,5.3 and 4.4  Mpc, respectively. These dwarf galaxies are characterized by high $ {\mbox{H\,{\sc i}}}$ mass content compared to the luminous mass and ${\mbox{H\,{\sc i}}}$ mass is comparable with other massive galaxies. The physical properties of these galaxy samples are presented in Table 1. The radial profile of both the stellar surface density in the B-band, kinematics as well as the gas surface density, were taken from \citeauthor{Begum_2008} (\citeyear{Begum_2008}), \citeauthor{Meurer_1996} (\citeyear{Meurer_1996}), \citeauthor{ooh_2015} (\citeyear{ooh_2015}) for NGC3741, NGC2915 and DDO168 respectively. All these input parameters are presented in Table 2

In Table \ref{input_parameters}, we present the input parameters used in DICE to simulate the sample galaxies. Simulation is run with $\mathrm{2\times10^6}$ dark matter halo particles, $3\times10^5$ gas disk particles and $\mathrm{3\times10^4}$ stellar particles following the methods outlined by {\citeauthor{WADSLEY2004137} (\citeyear{WADSLEY2004137}) and \citeauthor{Sellwood2019} (\citeyear{Sellwood2019}}. 
We used a Navarro-Frenk-White (NFW) dark matter halo for our simulations. This was preferred over a Pseudo-isothermal halo to comply with the observed shape of the rotation curve. The parameters like concentration and virial velocity were taken from the NFW mass models (NGC3741: \citeauthor{Begum_2008} \citeyear{Begum_2008}, NGC2915: \citeauthor{Meurer_1996} \citeyear{Meurer_1996}, DDO168: \citeauthor{ooh_2015} \citeyear{ooh_2015}). We run the simulations first with a spherical ($ c/a=1$), then oblate and, finally, with a prolate ($c/a > 1$) NFW dark matter halo with the spin parameter $\Lambda$ ranging from $ 0.01 - 0.07$.

\section{Results \& discussion}

\begin{table}
	\centering
    \label{tglobal_mode}
	\caption{Results from the global mode analysis for the gas disk of three galaxies}
	\label{table:2}
	\begin{tabular}{lccc} 
		\hline
		 Growth time (GYr) & NGC3741 & NGC2915 & DDO168\\
		\hline
		Stellar disk& 0.7&0.4&0.5 \\
		Gas disk &0.7  & 0.7 & 0.5 \\
        Dynamical time &0.1&0.15&0.12\\

		\hline
	\end{tabular}
\end{table}

\subsection{Global mode analysis}

Figure \ref{global_mode}  shows the global mode analysis for the $m=2$ mode in the stellar and gas disks of our galaxy sample. The stellar (gas) disk is assumed to respond to the net external potential of the gas (stars) and the dark matter halo. In the top panel, we present the eigenfrequency spectrum for the global m=2 mode in the stellar (Left Panel) and the gas (Right Panel) disk for NGC3741. The imaginary part of the eigenfrequency ($Im(\omega_i)$) gives the growth time of the corresponding mode in the disk, similarly, the middle and bottom panels for NGC2915 and DDO168, respectively. In Table 4, we present the growth time ($\frac{2 \pi}{Im(\omega_i)}$) for growth of m=2 mode in both stellar and gas disk. We note that the growth time ($\frac{2 \pi}{Im(\omega_i)}$) of the perturbation in both the stellar and the gas disk is a few dynamical times, thus implying that both the components are equally susceptible to the growth of $m=2$ global modes.
In Figure \ref{globalp}, we plot the amplitude of perturbation corresponding to the $m=2$ global mode at 1 Gyr. It clearly indicates peaks of amplitude 0.15, 0.2, and 0.3 within the bar radius for NGC 3741, NGC 2915, and DDO 168, respectively. This confirms the growth of $m=2$ perturbations and, hence, the formation of bars in the gas disks of the three dwarf galaxies.

\subsection{Simulations : N-body + Hydrodynamics}

\begin{figure}
    \centering
    \includegraphics[scale=0.5]{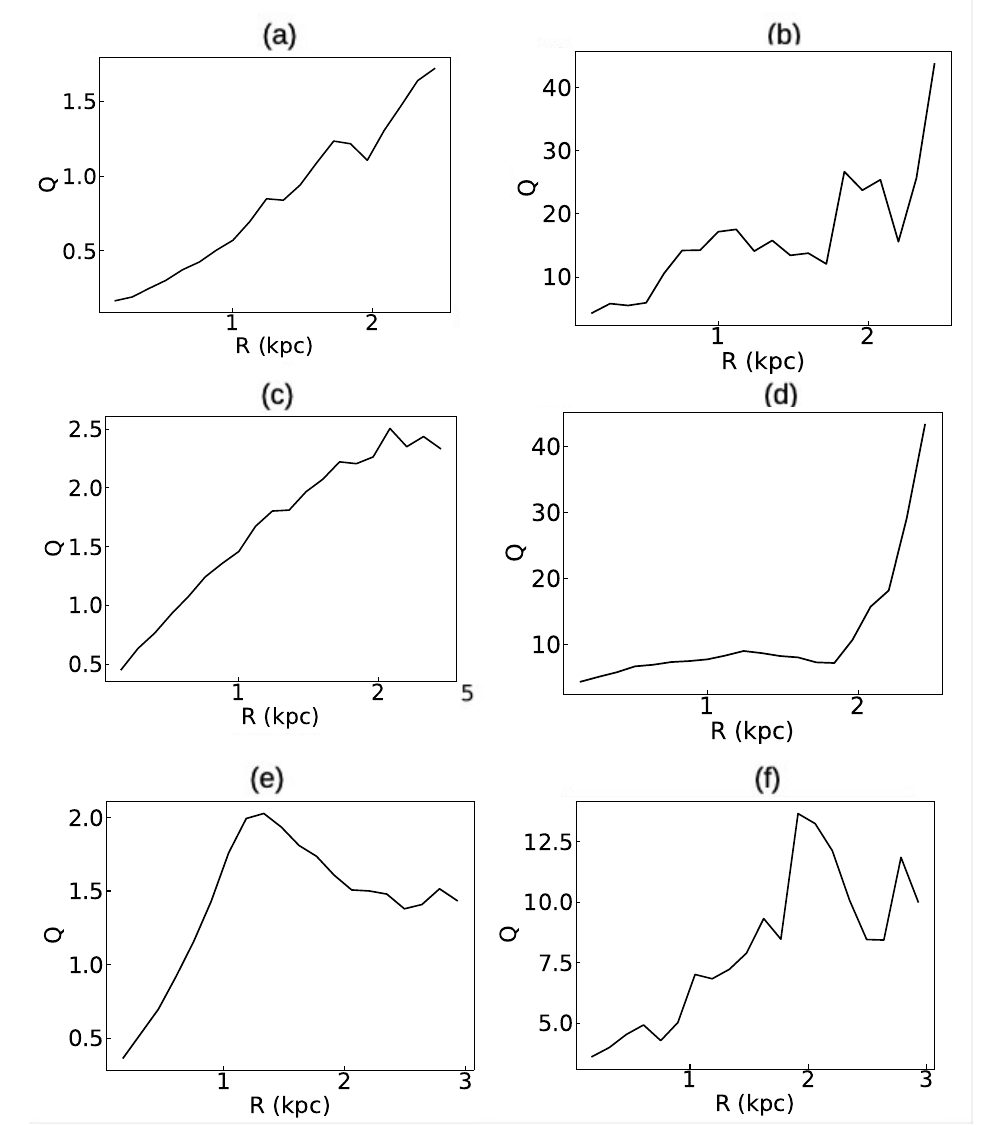}
    \caption{The radial profile of Toomre Q for the gas disk (Left) and the stellar disk (Right)  at the epoch which bar appeared in the gas disk. Top Row: NGC3741 Middle Row: NGC2915 Bottom Row: DDO168}
    \label{Toomre_Q}
\end{figure}

We use DICE to generate the initial conditions to simulate the purely gaseous bars in these dwarf galaxies. In Figure \ref{physical_paramater}, we confirm that our simulated galaxies comply with the observational constraints by matching the simulated rotation curves and radial gas surface density profiles with their observed counterparts in the initial conditions. We then feed the DICE output to RAMSES and evolve the galaxy for several epochs, about 12 dynamical times in each case. In Figure \ref{disk_snapshots_NGC3741}, in the top panel, we present the snapshots of the gas disk of NGC3741 at the time when the simulation started (Left) and when the bar and spiral arms just appeared (Right). Similarly, in the bottom panel, we show the same for the stellar disk. Figures \ref{disk_snapshots_NGC2915} and \ref{disk_snapshots_DDO168} show the same plots for NGC2915 and DDO168. We note that the bar instabilities appeared after 0.4, 0.5, and 0.3 Gyrs (corresponding to 4, 5, and 2 dynamical times) for NGC2915, NGC3741, and DDO168, respectively. In each case, the bar is accompanied with spiral arms and survives for 10 dynamical times. \\

To check if our results are robust against the choice of resolution of the simulation, we also perform a higher-resolution simulation of one of our sample galaxies, NGC2915, using $2 \times 10^7$ dark matter halo particles, $3 \times 10^6$ gas particles, and $3 \times 10^5$ stellar particles. Comparison of the radial profiles of the normalized amplitude of the $m=2$ mode between the low and high-resolution cases showed a difference of less than 0.05 in the 1–2 kpc range. Since the bar radii of our sample galaxies are less than 1 kpc, this difference is too small to change our conclusions regarding the presence or absence of a bar in our sample galaxies. Besides, comparing the rotation curves and gas surface density profiles from the low and high-resolution simulations, we find that those, too, are consistent within the error bars of the observations. Finally, comparing snapshots of the gaseous disks does not indicate any significant difference.

\noindent Interestingly, we observe a ring-shaped structure in the stellar disk of galaxy DDO168 at the epoch in which a bar is formed in the gas disk. Different mechanisms are involved in the formation of galactic rings. Resonant rings are formed from gas accumulation at certain resonances because of the continuous action of gravity torques produced by the bars; they can also be part of the spiral pattern (\citeauthor{Buta_combes} \citeyear{Buta_combes}, \citeauthor{Kormendy} \citeyear{Kormendy}). The resonant ring-shaped patterns are of two types: inner ring, which is associated with inner Lindblad resonance and outer rings, which are associated with outer Lindblad resonance \citeauthor{Buta_combes} \citeyear{Buta_combes}; here in our simulation, we observe an inner ring in the stellar disk. However, the existence of rings in unbarred galaxies also has been identified (\citeauthor{Buta_combes} \citeyear{Buta_combes}). \\

\begin{figure*}
\begin{center}
\begin{tabular}{cccc}

\resizebox{55mm}{!}{\includegraphics{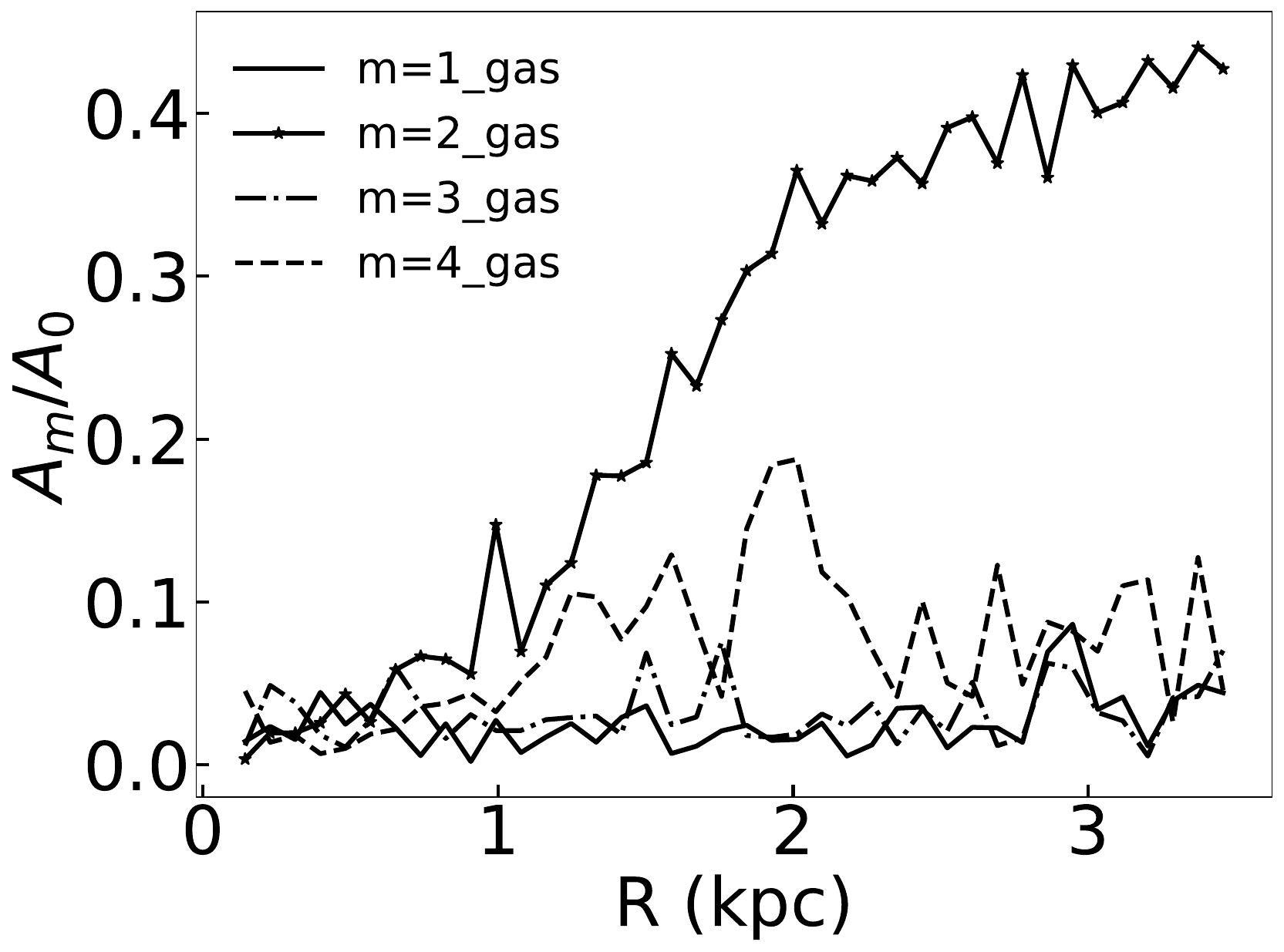}}\textbf{(a)}
\resizebox{55mm}{!}{\includegraphics{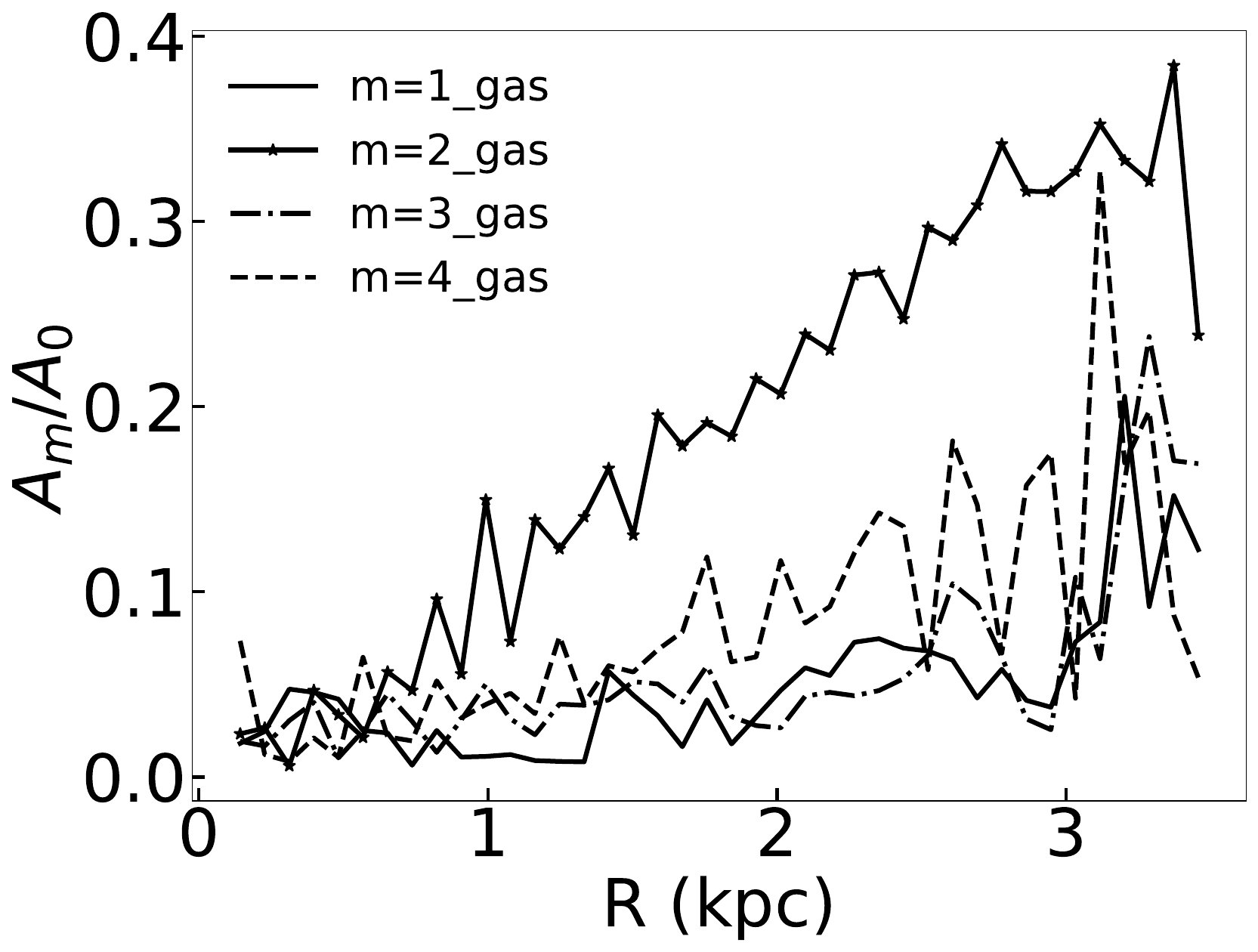}}\textbf{(b)}
\resizebox{55mm}{!}{\includegraphics{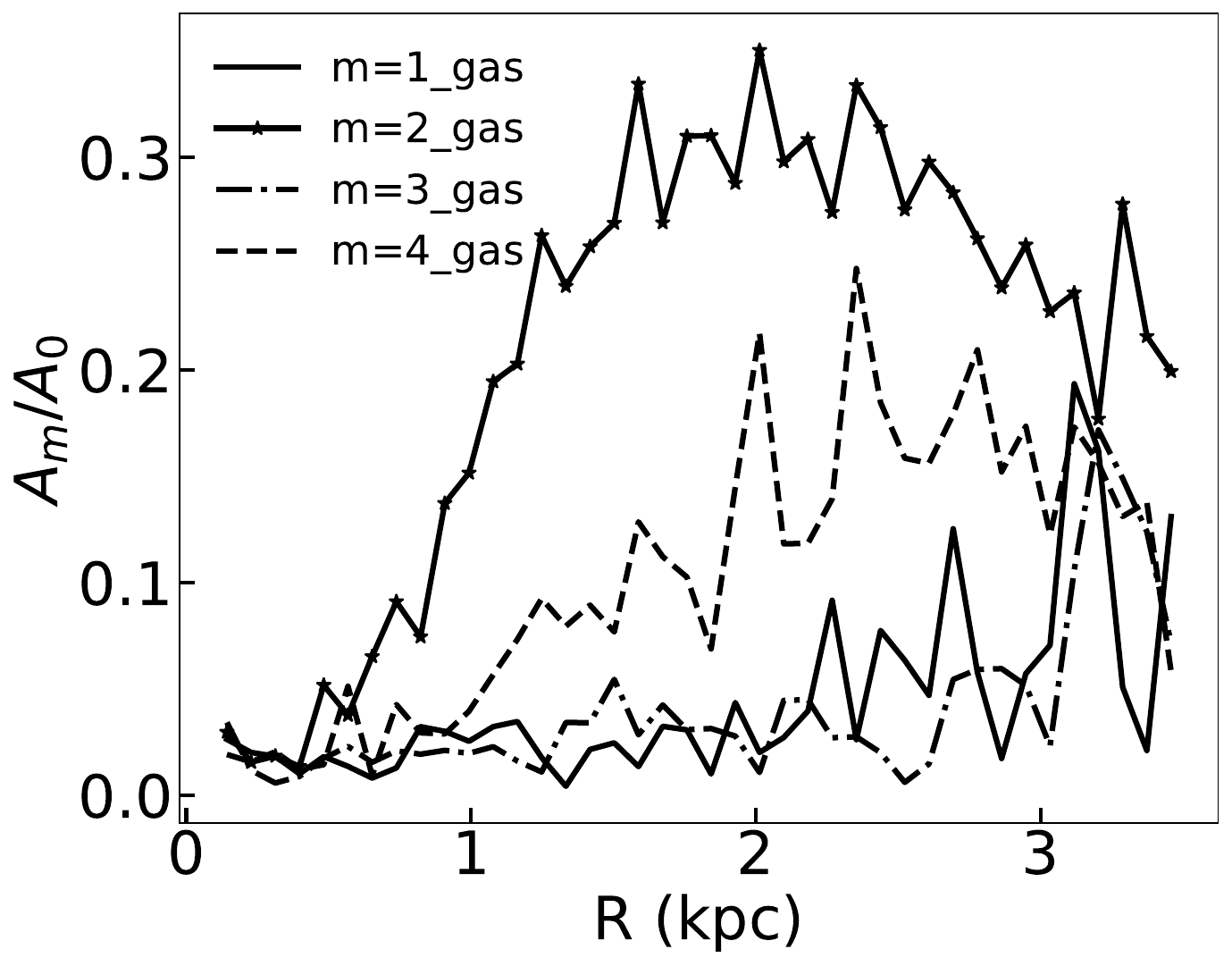}}\textbf{(c)}
\end{tabular}
\end{center}
\caption{Amplitudes of the different Fourier modes of the gas disk as a function of radius: (a)  NGC3741 (b) NGC2915 (c) DDO168 }
\label{fourier_amplitude}
\end{figure*}

\noindent \textbf{An oblate dark matter halo with a high spin parameter:} 
Interestingly, we find that the presence of an oblate dark-matter halo, with the planar-to-vertical axes ratio $c/a$ ranging between 0.6 to 0.8,  is necessary for the formation of a bar in the gas disk for all three galaxies. We confirm this by running the simulations first with a spherical ($c/a=1$), then oblate and, finally, with a prolate ($c/a > 1$) NFW dark matter halo, we observe that the bar structure in the gas disk appears only for an oblate dark-matter halo. Hence, the quadrupole moment of an underlying gravitational potential of the dark matter halo in the plane of the galactic disk is essential to drive the formation of bars and spiral arms in these dwarf galaxies. Interestingly, however, oblate to prolate shapes of dark matter halos form bars, but the bar is formed earlier in galaxies with prolate dark matter halos. Besides, Milky Way-type galaxies in cosmological zoom-in simulations show a triaxial halo with median b/a = 0.9 and c/a = 0.9 (\citeauthor{Prada_2019} \citeyear{Prada_2019}). Furthermore, a high value of the spin parameter of the dark matter halo $\Lambda = 0.04 - 0.07$ was essential for bar and spiral arm formation in our simulations. In cosmological hydrodynamical simulations like the Illustris TNG, the mean halo spin value is $0.035$ for barred galaxies (\citeauthor{Rosas-Guevara2021} \citeyear{Rosas-Guevara2021}). It has been found that dark-matter halos with spin parameters in the range of $0.01-0.07$ promote the formation of bars (\citeauthor{Weinberg_1985} \citeyear{Weinberg_1985}). Halos with high $\Lambda$ absorb the angular momentum from the disk more efficiently and enhance the growth of the bar (\citeauthor{Saha_2013} \citeyear{Saha_2013}). \\

\begin{figure}
    \centering
    \includegraphics[scale=0.3]{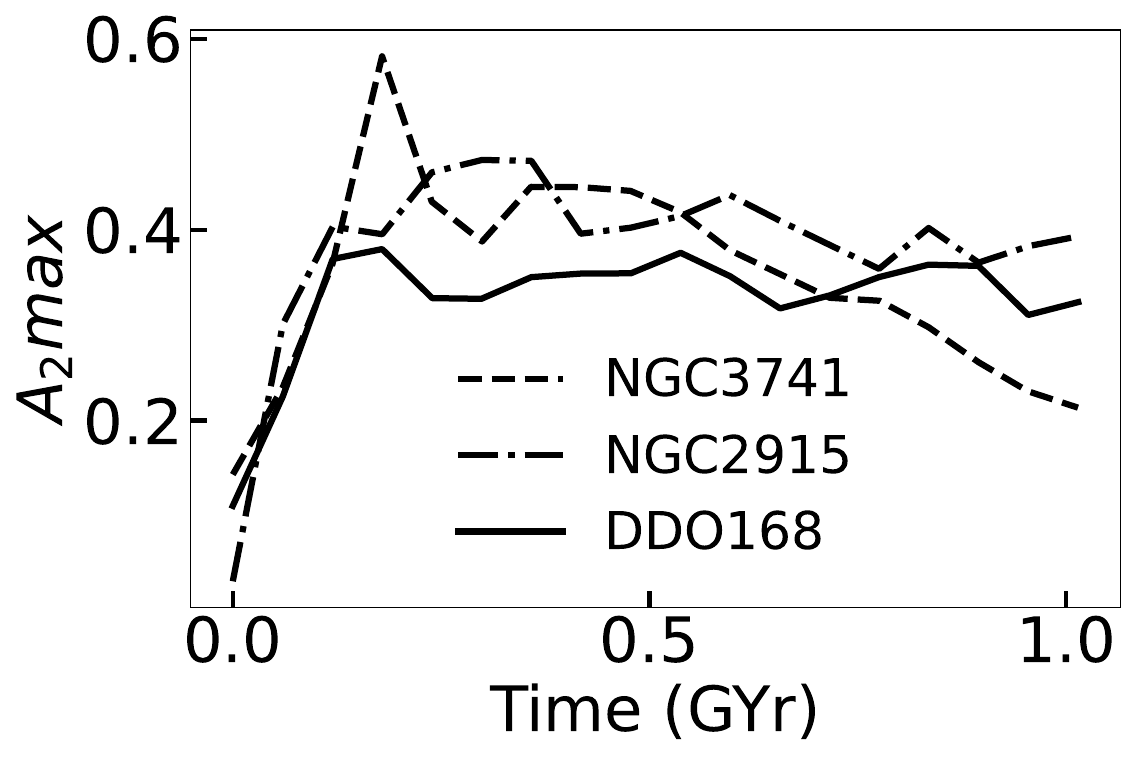}
    \caption{The maximum value of the amplitude of the $m=2$ mode as a function of time for the simulated galaxies (a)  NGC3741 (b) NGC2915 (c) DDO168}
    \label{bar_strength}
\end{figure}

\noindent Figure \ref{Toomre_Q} shows the radial profile of Toomre Q for the gas disk (left) and the stellar disk (right) at the epoch at which the bar appeared in the gas disk. Top row NGC3741, middle row NGC2915, and bottom row DDO168, respectively. For each galaxy, the stellar disk shows a higher Toomre Q value than the gas disk. \\

\noindent \textbf{Bar-Strength \& Pattern Speed:} 
We perform Fourier Analysis of the simulated gas surface density maps of the galaxies to obtain the amplitudes of the different modes, which indicate the strength of the various non-axisymmetric features. The $m=2$ mode is typically associated with a bar or a spiral arm. In Figure \ref{fourier_amplitude}, we present the first few Fourier modes as a function of galactocentric radius for NGC3741 (left), NGC2915 (middle), and DDO168 (right). The relatively high values of the amplitudes for the $m=2$ modes ($A_2$) for all three galaxies confirm the presence of a gaseous bar in their disks, as $m_2 > 0.25$ can be considered to be a strong bar (\citeauthor{Athanassoula1992;} \citeyear{Athanassoula1992;}, \citeauthor{Buta1993} \citeyear{Buta1993}).  
In addition, in Figure \ref{bar_strength}, we study the evolution of the maximum bar strength as a function of time for all our sample galaxies. For NGC2915 and DDO168, $A_2$ remains relatively constant around 0.3, while for NGC3741, $A_2$ falls from 0.4 to 0.3 and remains constant.

In N-body simulations, Fourier analysis or methods like Tremaine-Weinberg can estimate pattern speeds, but spatial variations and interactions between structures, like bars and spiral arms, suggest the need for a more adaptable approach (\citeauthor{Daniel2023} \citeyear{Daniel2023}).
Here, we obtain the pattern speed corresponding to the $m=2$ modes using both the Tremaine-Weinberg method and the Fourier transform. In the Tremaine-Weinberg method, the pattern speed obtained for all three galaxies fluctuated significantly with respect to the radius. Interestingly, the same technique gave constant pattern speed for these gaseous bars in the ${\mbox{H\,{\sc i}}}$ observation. We also calculated the pattern speed from the Fourier transform method. The average values of the pattern speeds are about -14.9, 9 and -16.7$\mathrm{km}~\mathrm{s}^{-1}\mathrm{kpc}^{-1}$ for NGC3741, NGC2915, and DDO168, respectively, at the epoch in which it matches those obtained from the ${\mbox{H\,{\sc i}}}$ observations mentioned in section \ref{samples}. However the average speed varies with time in the range  11.6 - 18.6 $\mathrm{km}~\mathrm{s}^{-1}\mathrm{kpc}^{-1}$,  3.3-14.9 $\mathrm{km}~\mathrm{s}^{-1}\mathrm{kpc}^{-1}$ and 13.4-20.3$\mathrm{km}~\mathrm{s}^{-1}\mathrm{kpc}^{-1}$ for NGC3741, NGC2915 and DDO168 respectively. This implies that the pattern seen is not a density wave. Therefore, we can conclude that the gaseous bars seen in these galaxies are not density waves but transient patterns lasting for more than ten dynamical times. \\ \\

\begin{table*}
    \centering
    \caption{Results from N-body + hydrodynamical simulation of our sample galaxies}
    \begin{tabular}{l|cc|cc|cc} 
        \hline
        \multirow{2}{*}{} & \multicolumn{2}{c|}{NGC 3741} & %
        \multicolumn{2}{c|}{NGC 2915} & \multicolumn{2}{c}{DDO 168} \\
        & Observed & Simulation & Observed & Simulation & Observed & Simulation  \\
        \hline
        Gaseous bar & Yes & Yes & Yes & Yes & Yes & Yes \\
        Magnitude of pattern speed ($~km~s^{-1}kpc^{-1}$) & 17.1 $\pm$ 3.4 & 14.9 & 8.0 $\pm$ 2.4 & 9 & 23.3 $\pm$ 5.9 & 16.7 \\
        Bar length ($kpc$) & 1.2 & 1.2 & 1.0 & 1.0 & 1.8 & 1.8 \\
        \hline
    \end{tabular}
    \label{results}
\end{table*}

\begin{figure*}
\begin{center}
    \includegraphics[scale=0.5]{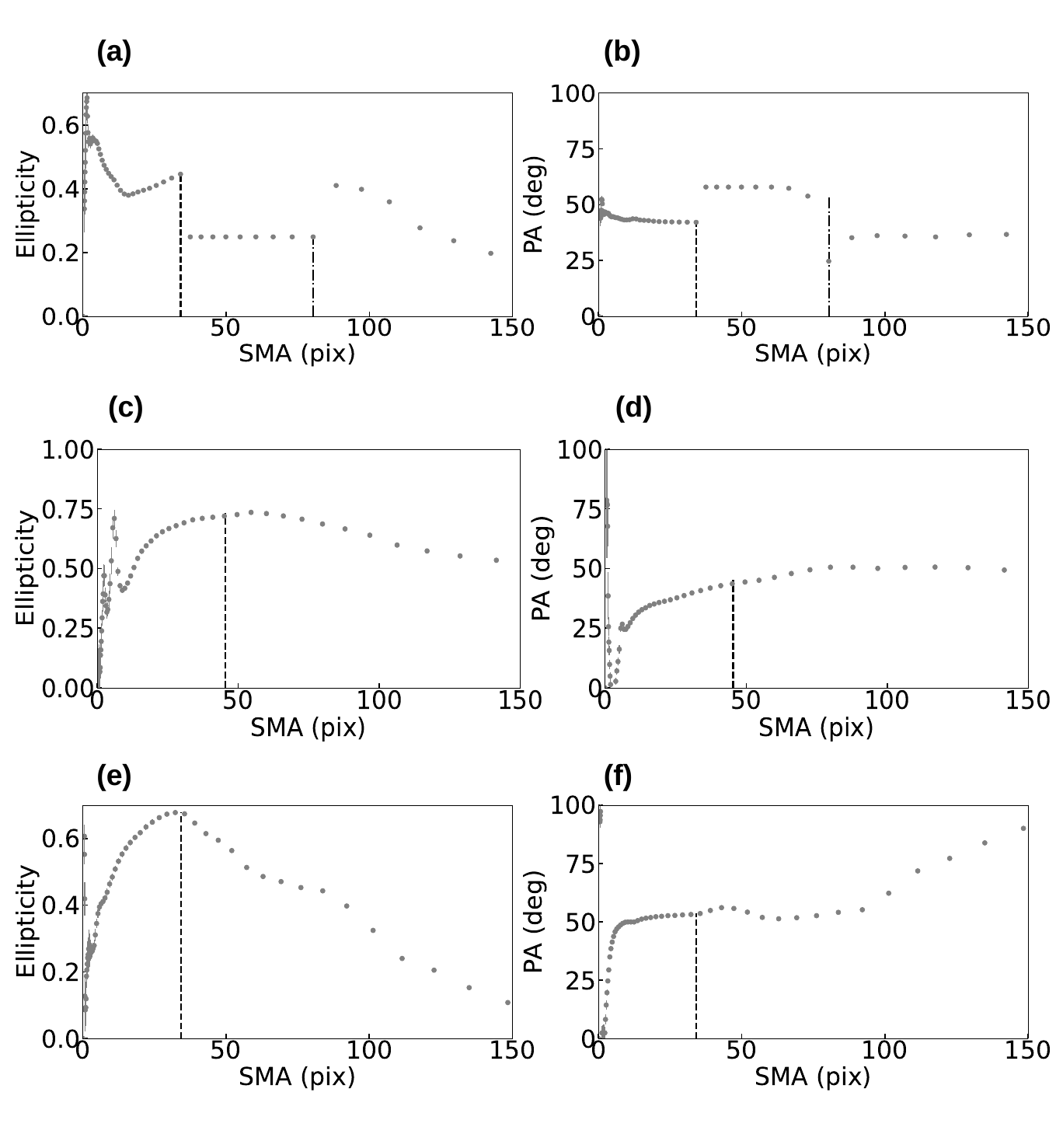}
\end{center}
\caption{Ellipticity of the iso-density contours (Left Panel) and Position Angle of the major axes of the same (Right Panel) of the simulated galaxies as a function of the galactocentric radius. The dotted vertical line shows the abrupt change in slope in each case  Top Row: NGC3741 Middle Row: NGC2915 Bottom Row: DDO168}
\label{ellipse_fit}
\end{figure*}

\begin{figure*}
    \centering
    \includegraphics[scale=0.37]{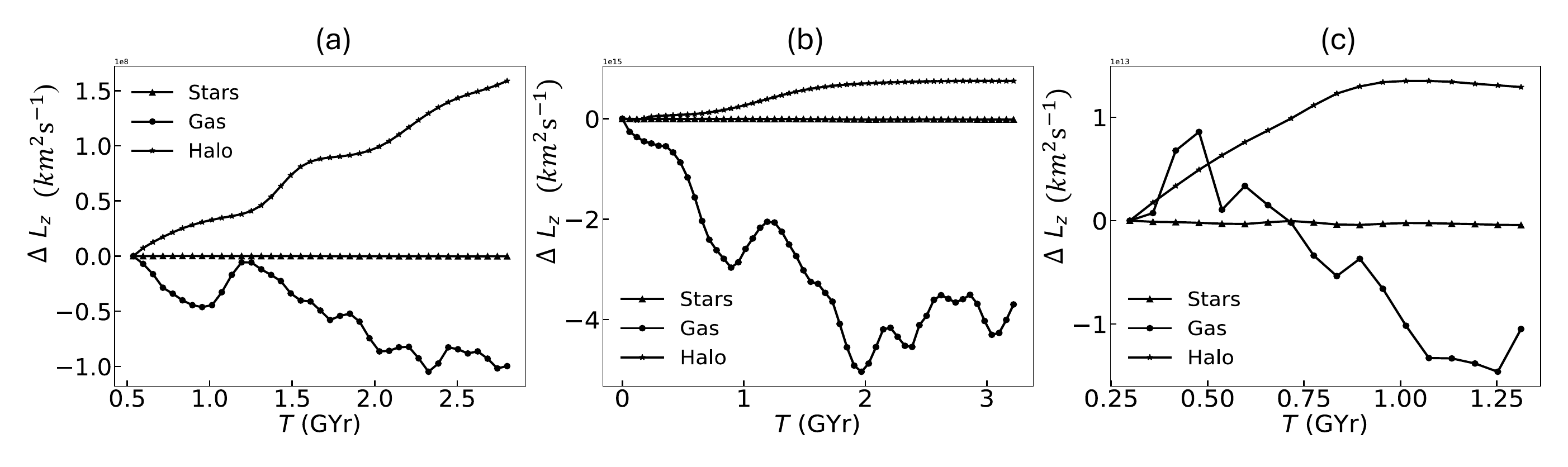}
    \caption{Change in angular momentum of the stars, the gas, and the dark matter halo as a function of time for the simulated galaxies: (a) NGC3741 (b) NGC2915 (c) DDO168}
    \label{Ang_mom}
\end{figure*}

\begin{figure*}
\begin{center}
\begin{tabular}{cccc}

\resizebox{55mm}{!}{\includegraphics{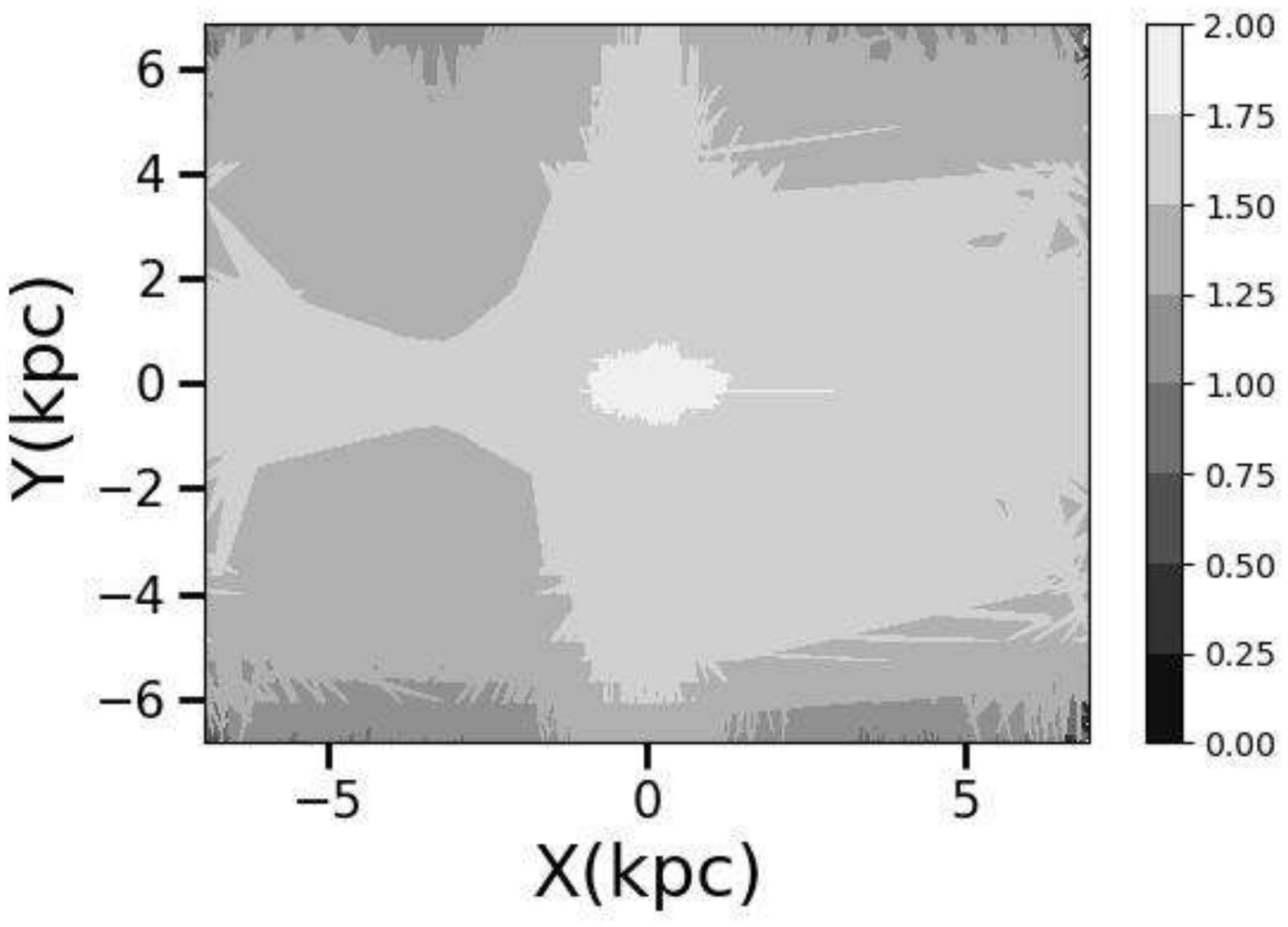}}\textbf{(a)}
\resizebox{55mm}{!}{\includegraphics{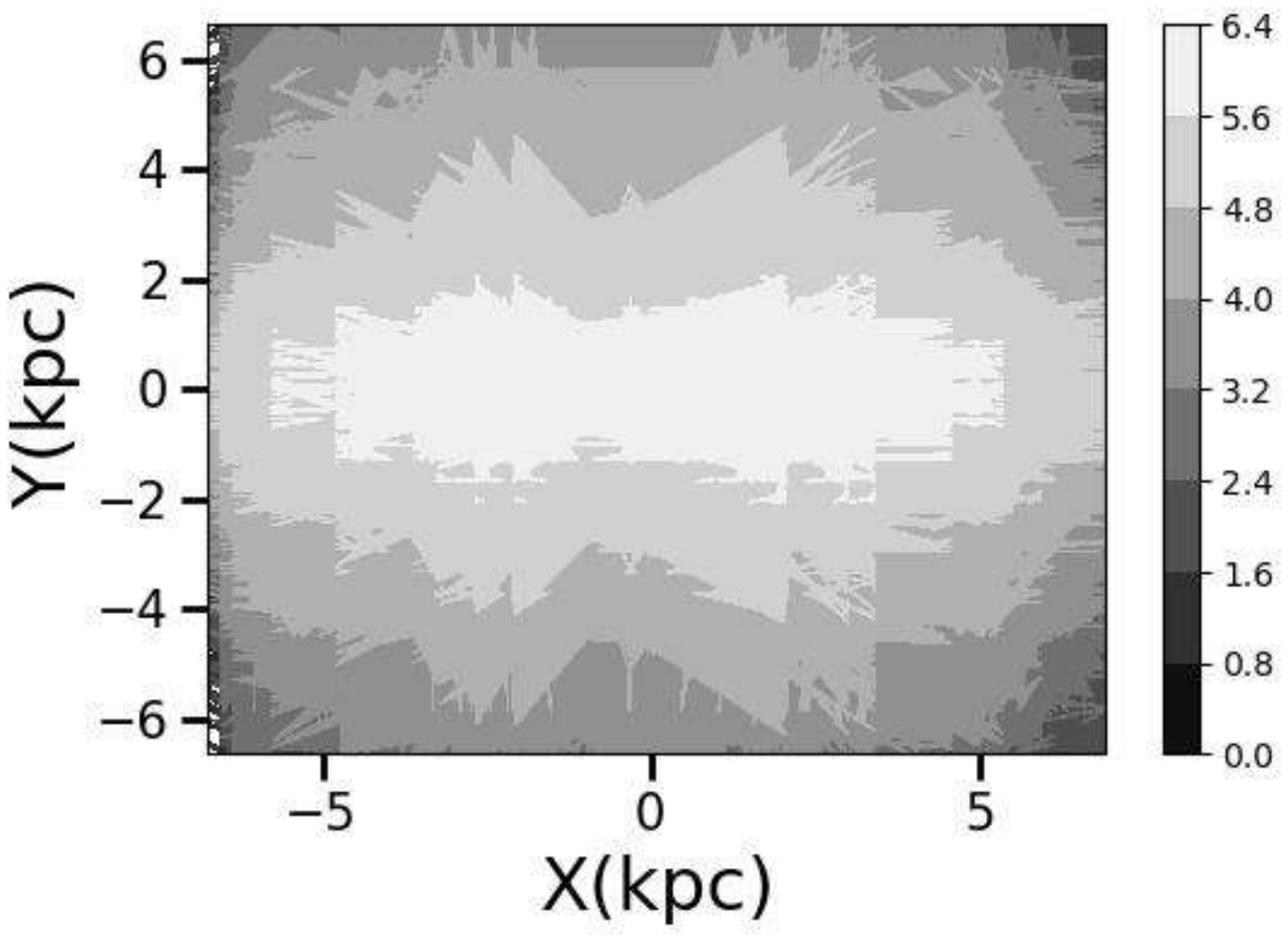}}\textbf{(b)}
\resizebox{55mm}{!}{\includegraphics{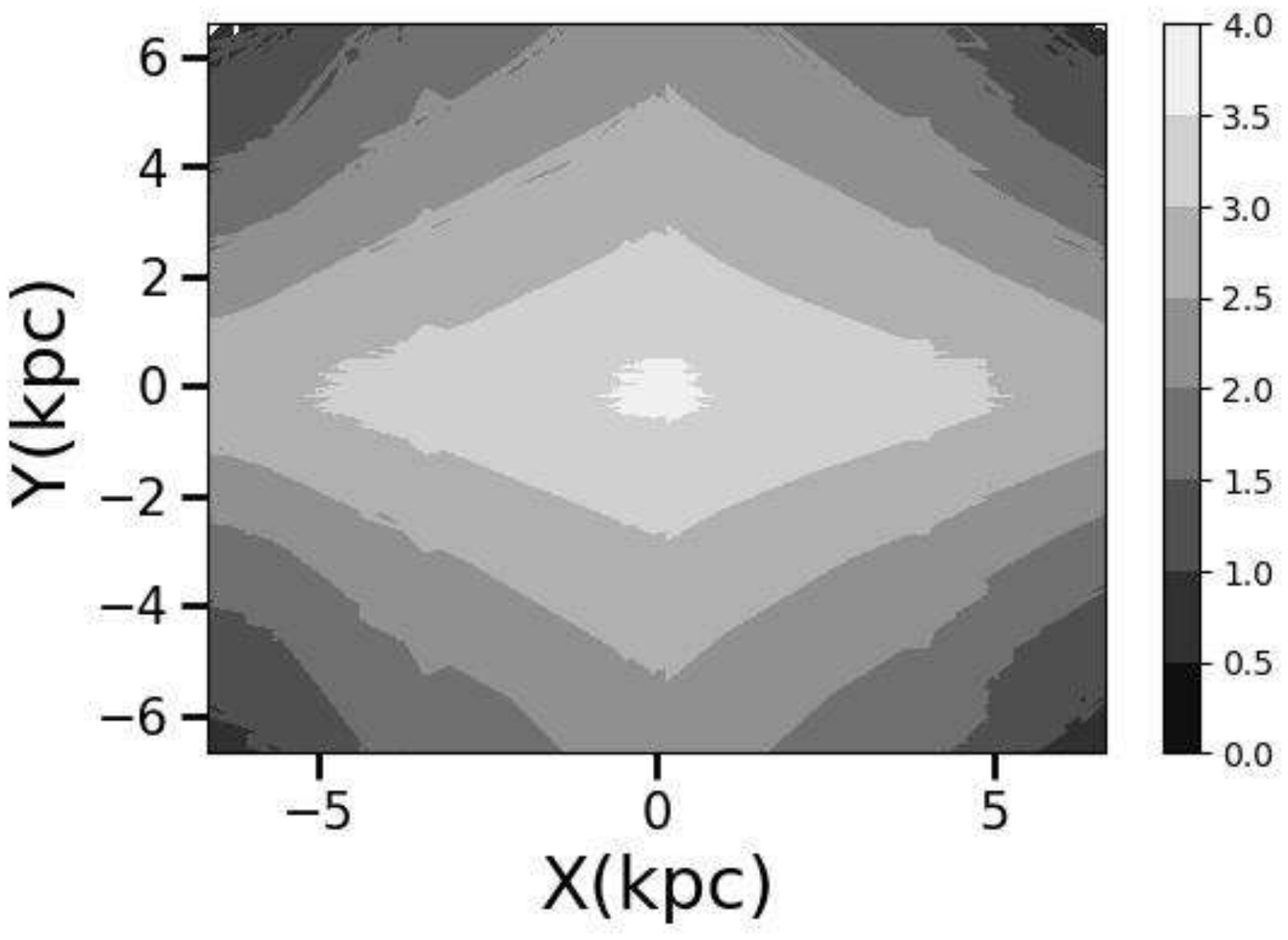}}\textbf{(c)}
\end{tabular}
\end{center}
\caption{Map of the Mach number over the gas disk of the simulated galaxies a) NGC3741, b)NGC2915, and c) DDO168. }
\label{shock}
\end{figure*}

\noindent  \textbf{Length and eccentricity of the bar:}
In Figure \ref{ellipse_fit}, we present the radial profiles showing the variation of (Left) ellipticity ($e$)  and (Right) position angle (PA) along the semi-major axis of the fitted ellipses for NGC3741 (Top Panel), NGC2915 (Middle Panel) and DDO168 (Bottom Panel). These were generated by fitting ellipses to the gas surface density profile using the python module \emph{isophote} (\citeauthor{larry_bradley_2022_6825092} \citeyear{larry_bradley_2022_6825092}). For NGC3741, we observe an abrupt change of 0.1 in the ellipticity profiles at 30" and 80";  the position angle changes by 10 degrees at the same radial points. For NGC2915, both the ellipticity and the position angle gradually increase up to 50" and then flatten out. Finally, for DDO168, the ellipticity and the position angle profiles both indicate the first change in slope change around 30" and the second one at 80". We measure the bar length as the semi-major axis length corresponding to the maximum ellipticity (\citeauthor{WP} \citeyear{WP}; \citeauthor{Menéndez-Delmestre_2007} \citeyear{Menéndez-Delmestre_2007}).
 \\

\noindent In Table \ref{results}, we present our calculated average pattern speeds for the $m=2$ mode and also the bar length for our simulated galaxies: 1.2 kpc (NGC3741), one kpc (NGC2915), and 0.8 kpc (DDO168) which matches well with the values determined from direct observations \citeauthor{Kam2017} \citeyear{Kam2017}; \citeauthor{Banerjee2013} \citeyear{Banerjee2013}; \citeauthor{PatraJog2019} \citeyear{PatraJog2019}). \\ \\

\noindent \textbf{Redistribution of angular momenta among disk components:} The nonspherical bar plays a crucial role in redistributing the angular momentum in galaxies. This dynamical process strengthens the bar over time, in which the bar grows, but its pattern speed is reduced (\citeauthor{Lynden-BellandKalnajs1972} \citeyear{Lynden-BellandKalnajs1972}), \citeauthor{Athanassoula2005} (\citeyear{Athanassoula2005}) and \citeauthor{Weinberg1985} (\citeyear{Weinberg1985}) showed that the bar transfers angular momentum to the halo, which in turn affects the growth of the bar, and the bar eventually slows down. Using angular momentum exchange between disk and halo, \citeauthor{Kataria_2022} (\citeyear{Kataria_2022}), \citeauthor{Collier_2019} (\citeyear{Collier_2019}) studied the role of halo spin in bar formation and buckling  \citeauthor{frosst2024} \citeyear{frosst2024} showed that the angular momentum exchange follows the bar formation phases, increasing during assembly, briefly leveling off during buckling, and rising again near a steady state. These angular momentum transfers occur in coevolving halos (\citeauthor{Sellwood_2008} (\citeyear{Sellwood_2008}); in the live halo model, angular momentum transfer continues, while in a static halo, this transfer halts abruptly. 

In Figure \ref{Ang_mom}, we plot the change in the angular momentum of the stellar, gas, and dark matter components after the bar appears in the disk for each of our sample galaxies. We observe a noticeable decrease in the angular momentum of the gas disk and a corresponding increase for the dark-matter halo, thus indicating an angular momentum exchange between these two components. However, there is no discernible change in the angular momentum of the stellar disk, underscoring that the stellar bar formed is weak. \\

\noindent \textbf{Survival of a gaseous bar in the presence of shock waves:}
\noindent The non-axisymmetric quadrupole moment of the potential associated with the bar can induce shocks which may rip off the self-gravitating bar    (\citeauthor{Sorensen1976} \citeyear{Sorensen1976}, \citeauthor{Roberts1979} \citeyear{Roberts1979}, \citeauthor{Kovalenko1999} \citeyear{Kovalenko1999}, \citeauthor{Azam2022} \citeyear{Azam2022}). To assess the strength and significance of these shocks, we plotted the local Mach number ($M=(|v|/c)$, where $(|v|)$  represents the magnitude of the gas velocity and c is the local sound speed) in the gas disks in Figure 11: (Left) NGC3741, (Middle) NGC2915, and (Right) DDO168. The maximum values of the Mach number within the gas disks of these galaxies are as follows: 2 (NGC3741), 6 (NGC2915), and 4 (DDO168). Since the Mach numbers are relatively low, the shock waves may not be strong enough to disrupt the gaseous bar, and it survives. This is unlike the scenario seen in some low surface brightness galaxies F568-VI and F568-01, where the 
Mach number can be in the range of 20-40, and no non-axisymmetric features present in the gas disk (\citeauthor{ganesh_2024} \citeyear{ganesh_2024}).

 Dwarf galaxies are generally characterized by low star formation rates (SFR) (\citeauthor{Hunter_1985} \citeyear{Hunter_1985}), and so are our sample dwarf irregulars. For example, the SFR of NGC3741, NGC2915 and  DDO168 are $4.3\times10^{-3}$ (\citeauthor{Begum_2008} \citeyear{Begum_2008}), $5 \times 10^{-3}$ (\citeauthor{Meurer_1996} \citeyear{Meurer_1996}), and $6.4 \times 10^{-3}$ (\citeauthor{Hunter_2004} \citeyear{Hunter_2004}). We use the SFR value as an input parameter in our simulations. Further, it is now well-known that the evolution of galactic bars may be significantly impacted by SN feedback (\citeauthor{bar_feedback} \citeyear{bar_feedback}). To investigate this, we re-ran the simulations with a non-zero $\eta$ where $\eta$ is the mass fraction of newly formed stars that explode into a supernova and hence capture the effect of supernova feedback. We chose $\eta=0.05$ following \citeauthor{ganesh_2024} (\citeyear{ganesh_2024}) who modeled observed low surface brightness galaxies using N-body + hydrodynamic simulations. Interestingly, neither the Mach Number plots nor the bar amplitude change even after including the effects of feedback, possibly due to the very low mass fraction of the stars in dwarf galaxies.

\section{Conclusions}
We construct dynamical models of a purely gaseous bar, i.e., possibly one without a stellar counterpart, as observed in three dwarf galaxies: NGC3741, NGC2915, and DDO168, as constrained by mass models available in the literature. We first model the gas (stellar) disk as a self-gravitating component, also subjected to the external potential of the stars (gas) and a spherical dark matter halo, and study its response to global $m=2$ instabilities. We note that our gas and stellar disks are moderately unstable against the growth of global $m=2$ modes. Finally, we use the publicly available N-body + hydrodynamical simulation code RAMSES to study the formation of purely gaseous bars in these galaxies. Our study shows that an oblate halo with axis ratio $c/a=0.7$ and a high value of spin parameter $\Lambda = 0.04 - 0.07$ forms a gaseous bar, which survives for more than ten dynamical times. Our calculated Mach number $M=4 - 6$ confirms that the shock waves developed are not strong enough to rip off the gaseous bars. Although observational studies have not detected the presence of a stellar bar in these galaxies, our simulations show the formation of a tiny bar in each case. However, the evolution of the change in angular momentum $L_z$ of the different disk components indicates that the stellar bar, even if present, is weak.

\section*{Acknowledgements}
The authors thank Dr. Ganesh Narayanan, IISER Tirupati for sharing the numerical codes for the global mode study. They also thank Prof. Bhargav Vaidya, Indian Institute of Technology, Indore and the anonymous referee for valuable comments and suggestions.

\bibliographystyle{mnras}
\bibliography{references} 

\begin{thebibliography}{}
\makeatletter
\relax
\def\mn@urlcharsother{\let\do\@makeother \do\$\do\&\do\#\do\^\do\_\do\%\do\~}
\def\mn@doi{\begingroup\mn@urlcharsother \@ifnextchar [ {\mn@doi@} {\mn@doi@[]}}
\def\mn@doi@[#1]#2{\def\@tempa{#1}\ifx\@tempa\@empty \href {http://dx.doi.org/#2} {doi:#2}\else \href {http://dx.doi.org/#2} {#1}\fi \endgroup}
\def\mn@eprint#1#2{\mn@eprint@#1:#2::\@nil}
\def\mn@eprint@arXiv#1{\href {http://arxiv.org/abs/#1} {{\tt arXiv:#1}}}
\def\mn@eprint@dblp#1{\href {http://dblp.uni-trier.de/rec/bibtex/#1.xml} {dblp:#1}}
\def\mn@eprint@#1:#2:#3:#4\@nil{\def\@tempa {#1}\def\@tempb {#2}\def\@tempc {#3}\ifx \@tempc \@empty \let \@tempc \@tempb \let \@tempb \@tempa \fi \ifx \@tempb \@empty \def\@tempb {arXiv}\fi \@ifundefined {mn@eprint@\@tempb}{\@tempb:\@tempc}{\expandafter \expandafter \csname mn@eprint@\@tempb\endcsname \expandafter{\@tempc}}}

\bibitem[\protect\citeauthoryear{{Aguerri}}{{Aguerri}}{1999}]{Aguerri1999}
{Aguerri} J.~A.~L.,  1999, \aap, \href {https://ui.adsabs.harvard.edu/abs/1999A&A...351...43A} {351, 43}

\bibitem[\protect\citeauthoryear{{Aguerri}, {M{\'e}ndez-Abreu}  \& {Corsini}}{{Aguerri} et~al.}{2009}]{Aguerrietal2009}
{Aguerri} J.~A.~L.,  {M{\'e}ndez-Abreu} J.,   {Corsini} E.~M.,  2009, \mn@doi [\aap] {10.1051/0004-6361:200810931}, \href {https://ui.adsabs.harvard.edu/abs/2009A&A...495..491A} {495, 491}

\bibitem[\protect\citeauthoryear{{Aguerri} et~al.,}{{Aguerri} et~al.}{2015}]{(Aguerrietal2015}
{Aguerri} J.~A.~L.,  et~al., 2015, \mn@doi [\aap] {10.1051/0004-6361/201423383}, \href {https://ui.adsabs.harvard.edu/abs/2015A&A...576A.102A} {576, A102}

\bibitem[\protect\citeauthoryear{{Athanassoula}}{{Athanassoula}}{1992}]{Athanassoula1992;}
{Athanassoula} E.,  1992, \mn@doi [\mnras] {10.1093/mnras/259.2.345}, \href {https://ui.adsabs.harvard.edu/abs/1992MNRAS.259..345A} {259, 345}

\bibitem[\protect\citeauthoryear{Athanassoula}{Athanassoula}{2005}]{Athanassoula2005}
Athanassoula E.,  2005, \mn@doi [Monthly Notices of the Royal Astronomical Society] {10.1111/j.1365-2966.2005.08872.x}, 358, 1477

\bibitem[\protect\citeauthoryear{{Azam}, {Sami}  \& {Rahaman}}{{Azam} et~al.}{2022}]{Azam2022}
{Azam} M.,  {Sami} M.,   {Rahaman} F.,  2022, \mn@doi [arXiv e-prints] {10.48550/arXiv.2202.02229}, \href {https://ui.adsabs.harvard.edu/abs/2022arXiv220202229A} {p. arXiv:2202.02229}

\bibitem[\protect\citeauthoryear{Banerjee, Patra, Chengalur  \& Begum}{Banerjee et~al.}{2013}]{Banerjee2013}
Banerjee A.,  Patra N.,  Chengalur J.,   Begum A.,  2013, \mn@doi [Monthly Notices of the Royal Astronomical Society] {10.1093/mnras/stt1083}, 434

\bibitem[\protect\citeauthoryear{Begum, Chengalur, Kennicutt, Karachentsev  \& Lee}{Begum et~al.}{2007}]{Begum_2008}
Begum A.,  Chengalur J.~N.,  Kennicutt R.~C.,  Karachentsev I.~D.,   Lee J.~C.,  2007, \mn@doi [Monthly Notices of the Royal Astronomical Society] {10.1111/j.1365-2966.2007.12592.x}, 383, 809

\bibitem[\protect\citeauthoryear{Bi, Shlosman  \& Romano}{Bi et~al.}{2022}]{bar_feedback}
Bi D.,  Shlosman I.,   Romano E.,  2022, \mn@doi [Monthly Notices of the Royal Astronomical Society] {10.1093/mnras/stac363}, 513, 693

\bibitem[\protect\citeauthoryear{Binney \& Tremaine}{Binney \& Tremaine}{1987}]{BinneyTremaine1987}
Binney J.,  Tremaine S.,  1987, pp 149--180

\bibitem[\protect\citeauthoryear{{Binney} \& {Tremaine}}{{Binney} \& {Tremaine}}{2008}]{Binney2008}
{Binney} J.,  {Tremaine} S.,  2008, {Galactic Dynamics: Second Edition}

\bibitem[\protect\citeauthoryear{{Peebles}}{Bod}{2007}]{Bodenheimeretal2007}
 2007, {Numerical Methods in Astrophysics: An Introduction}

\bibitem[\protect\citeauthoryear{{Bradley} et~al.,}{{Bradley} et~al.}{2020}]{Bradleyetal2020}
{Bradley} L.,  et~al., 2020, {astropy/photutils: 1.0.1}, Zenodo, \mn@doi{10.5281/zenodo.4049061}

\bibitem[\protect\citeauthoryear{Bradley et~al.,}{Bradley et~al.}{2022}]{larry_bradley_2022_6825092}
Bradley L.,  et~al., 2022, ] {10.5281/zenodo.6825092}

\bibitem[\protect\citeauthoryear{Bureau, Freeman, Pfitzner  \& Meurer}{Bureau et~al.}{1999}]{Bureau1999}
Bureau M.,  Freeman K.,  Pfitzner D.,   Meurer G.,  1999, \mn@doi [AJ] {10.1086/301064}, 118

\bibitem[\protect\citeauthoryear{{Buta} \& {Combes}}{{Buta} \& {Combes}}{1996}]{Buta_combes}
{Buta} R.,  {Combes} F.,  1996, \fcp, \href {https://ui.adsabs.harvard.edu/abs/1996FCPh...17...95B} {17, 95}

\bibitem[\protect\citeauthoryear{{Buta} \& {Crocker}}{{Buta} \& {Crocker}}{1993}]{Buta1993}
{Buta} R.,  {Crocker} D.~A.,  1993, \mn@doi [\aj] {10.1086/116514}, \href {https://ui.adsabs.harvard.edu/abs/1993AJ....105.1344B} {105, 1344}

\bibitem[\protect\citeauthoryear{{Buta}, {Alpert}, {Cobb}, {Crocker}  \& {Purcell}}{{Buta} et~al.}{1998}]{Butaetal1998}
{Buta} R.,  {Alpert} A.~J.,  {Cobb} M.~L.,  {Crocker} D.~A.,   {Purcell} G.~B.,  1998, \mn@doi [\aj] {10.1086/300494}, \href {https://ui.adsabs.harvard.edu/abs/1998AJ....116.1142B} {116, 1142}

\bibitem[\protect\citeauthoryear{{Buta} et~al.,}{{Buta} et~al.}{2015}]{Butaetal2015)}
{Buta} R.~J.,  et~al., 2015, \mn@doi [\apjs] {10.1088/0067-0049/217/2/32}, \href {https://ui.adsabs.harvard.edu/abs/2015ApJS..217...32B} {217, 32}

\bibitem[\protect\citeauthoryear{Collier, Shlosman  \& Heller}{Collier et~al.}{2019}]{Collier_2019}
Collier A.,  Shlosman I.,   Heller C.,  2019, \mn@doi [Monthly Notices of the Royal Astronomical Society] {10.1093/mnras/stz2144}, 488, 5788–5801

\bibitem[\protect\citeauthoryear{{Contopoulos} \& {Grosbol}}{{Contopoulos} \& {Grosbol}}{1989}]{ContopoulosandGrosbol1989;}
{Contopoulos} G.,  {Grosbol} P.,  1989, \mn@doi [\aapr] {10.1007/BF00873080}, \href {https://ui.adsabs.harvard.edu/abs/1989A&ARv...1..261C} {1, 261}

\bibitem[\protect\citeauthoryear{{Eskridge} et~al.,}{{Eskridge} et~al.}{2000}]{Eskridgeetal2000}
{Eskridge} P.~B.,  et~al., 2000, \mn@doi [\aj] {10.1086/301203}, \href {https://ui.adsabs.harvard.edu/abs/2000AJ....119..536E} {119, 536}

\bibitem[\protect\citeauthoryear{Fathi, Beckman, Piñol-Ferrer, Hernandez, Martínez-Valpuesta  \& Carignan}{Fathi et~al.}{2009}]{Fathi_2009}
Fathi K.,  Beckman J.~E.,  Piñol-Ferrer N.,  Hernandez O.,  Martínez-Valpuesta I.,   Carignan C.,  2009, \mn@doi [The Astrophysical Journal] {10.1088/0004-637X/704/2/1657}, 704, 1657

\bibitem[\protect\citeauthoryear{{Freeman}}{{Freeman}}{1970}]{Freeman1970}
{Freeman} K.~C.,  1970, \mn@doi [\apj] {10.1086/150474}, \href {https://ui.adsabs.harvard.edu/abs/1970ApJ...160..811F} {160, 811}

\bibitem[\protect\citeauthoryear{{Friedli} \& {Martinet}}{{Friedli} \& {Martinet}}{1993}]{MartinetFriedli1997}
{Friedli} D.,  {Martinet} L.,  1993, \aap, \href {https://ui.adsabs.harvard.edu/abs/1993A&A...277...27F} {277, 27}

\bibitem[\protect\citeauthoryear{Frosst, Obreschkow  \& Ludlow}{Frosst et~al.}{2024}]{frosst2024}
Frosst M.,  Obreschkow D.,   Ludlow A.~D.,  2024, The active role of co-evolving haloes in stellar bar formation (\mn@eprint {arXiv} {2408.03375}), \url {https://arxiv.org/abs/2408.03375}

\bibitem[\protect\citeauthoryear{Garland et~al.,}{Garland et~al.}{2024}]{garland2024}
Garland I.~L.,  et~al., 2024, Galaxy Zoo DESI: large-scale bars as a secular mechanism for triggering AGN (\mn@eprint {arXiv} {2406.20096}), \url {https://arxiv.org/abs/2406.20096}

\bibitem[\protect\citeauthoryear{{Guo} et~al.,}{{Guo} et~al.}{2023}]{Guoetal2023}
{Guo} Y.,  et~al., 2023, \mn@doi [\apjl] {10.3847/2041-8213/acacfb}, \href {https://ui.adsabs.harvard.edu/abs/2023ApJ...945L..10G} {945, L10}

\bibitem[\protect\citeauthoryear{Hunter \& Elmegreen}{Hunter \& Elmegreen}{2004}]{Hunter_2004}
Hunter D.~A.,  Elmegreen B.~G.,  2004, \mn@doi [The Astronomical Journal] {10.1086/424615}, 128, 2170

\bibitem[\protect\citeauthoryear{{Hunter} \& {Gallagher}}{{Hunter} \& {Gallagher}}{1985}]{Hunter_1985}
{Hunter} D.~A.,  {Gallagher} III J.~S.,  1985, \mn@doi [\apjs] {10.1086/191051}, \href {https://ui.adsabs.harvard.edu/abs/1985ApJS...58..533H} {58, 533}

\bibitem[\protect\citeauthoryear{{Jungwiert, B.}, {Combes, F.}  \& {Axon, D. J.}}{{Jungwiert, B.} et~al.}{1997}]{jung_1997}
{Jungwiert, B.} {Combes, F.}  {Axon, D. J.} 1997, \mn@doi [Astron. Astrophys. Suppl. Ser.] {10.1051/aas:1997236}, 125, 479

\bibitem[\protect\citeauthoryear{{Pfenniger}, {Saha}  \& {Wu}}{Kam}{2017}]{Kam2017}
 2017, ] {10.3847/1538-3881/aa79f3}, 154, 41

\bibitem[\protect\citeauthoryear{Kataria \& Shen}{Kataria \& Shen}{2022}]{Kataria_2022}
Kataria S.~K.,  Shen J.,  2022, \mn@doi [The Astrophysical Journal] {10.3847/1538-4357/ac9df1}, 940, 175

\bibitem[\protect\citeauthoryear{Kataria \& Vivek}{Kataria \& Vivek}{2023}]{kataria_2024}
Kataria S.~K.,  Vivek M.,  2023, \mn@doi [Monthly Notices of the Royal Astronomical Society] {10.1093/mnras/stad3383}, 527, 3366

\bibitem[\protect\citeauthoryear{Korchagin, Kikuchi, Miyama, Orlova  \& Peterson}{Korchagin et~al.}{2000}]{global}
Korchagin V.,  Kikuchi N.,  Miyama S.~M.,  Orlova N.,   Peterson B.~A.,  2000, \mn@doi [The Astrophysical Journal] {10.1086/309447}, 541, 565

\bibitem[\protect\citeauthoryear{{Kormendy}}{{Kormendy}}{1979}]{Kormendy}
{Kormendy} J.,  1979, \mn@doi [\apj] {10.1086/156782}, \href {https://ui.adsabs.harvard.edu/abs/1979ApJ...227..714K} {227, 714}

\bibitem[\protect\citeauthoryear{{Kormendy} \& {Kennicutt}}{{Kormendy} \& {Kennicutt}}{2004}]{KormendyandKennicutt2004}
{Kormendy} J.,  {Kennicutt} Robert~C. J.,  2004, \mn@doi [\araa] {10.1146/annurev.astro.42.053102.134024}, \href {https://ui.adsabs.harvard.edu/abs/2004ARA&A..42..603K} {42, 603}

\bibitem[\protect\citeauthoryear{{Kovalenko} \& {Lukin}}{{Kovalenko} \& {Lukin}}{1999}]{Kovalenko1999}
{Kovalenko} I.~G.,  {Lukin} D.~V.,  1999, Astronomy Letters, \href {https://ui.adsabs.harvard.edu/abs/1999AstL...25..215K} {25, 215}

\bibitem[\protect\citeauthoryear{{Laine}, {Shlosman}, {Knapen}  \& {Peletier}}{{Laine} et~al.}{2002}]{laine_2002}
{Laine} S.,  {Shlosman} I.,  {Knapen} J.~H.,   {Peletier} R.~F.,  2002, \mn@doi [\apj] {10.1086/323964}, \href {https://ui.adsabs.harvard.edu/abs/2002ApJ...567...97L} {567, 97}

\bibitem[\protect\citeauthoryear{{Lynden-Bell} \& {Kalnajs}}{{Lynden-Bell} \& {Kalnajs}}{1972a}]{LyndenBellKalnajs1972}
{Lynden-Bell} D.,  {Kalnajs} A.~J.,  1972a, \mn@doi [\mnras] {10.1093/mnras/157.1.1}, \href {https://ui.adsabs.harvard.edu/abs/1972MNRAS.157....1L} {157, 1}

\bibitem[\protect\citeauthoryear{Lynden-Bell \& Kalnajs}{Lynden-Bell \& Kalnajs}{1972b}]{Lynden-BellandKalnajs1972}
Lynden-Bell D.,  Kalnajs A.~J.,  1972b, \mn@doi [Monthly Notices of the Royal Astronomical Society] {10.1093/mnras/157.1.1}, 157, 1

\bibitem[\protect\citeauthoryear{{Marinova} \& {Jogee}}{{Marinova} \& {Jogee}}{2007}]{MarinovaandJogee2007}
{Marinova} I.,  {Jogee} S.,  2007, \mn@doi [\apj] {10.1086/512355}, \href {https://ui.adsabs.harvard.edu/abs/2007ApJ...659.1176M} {659, 1176}

\bibitem[\protect\citeauthoryear{{Masset} \& {Bureau}}{{Masset} \& {Bureau}}{2004}]{Masset_2004}
{Masset} F.~S.,  {Bureau} M.,  2004, in {Ryder} S.,  {Pisano} D.,  {Walker} M.,   {Freeman} K.,  eds,  Vol. 220, Dark Matter in Galaxies. p.~293

\bibitem[\protect\citeauthoryear{Menéndez-Delmestre, Sheth, Schinnerer, Jarrett  \& Scoville}{Menéndez-Delmestre et~al.}{2007}]{Menéndez-Delmestre_2007}
Menéndez-Delmestre K.,  Sheth K.,  Schinnerer E.,  Jarrett T.~H.,   Scoville N.~Z.,  2007, \mn@doi [The Astrophysical Journal] {10.1086/511025}, 657, 790

\bibitem[\protect\citeauthoryear{{Meurer}, {Carignan}, {Beaulieu}  \& {Freeman}}{{Meurer} et~al.}{1996}]{Meurer_1996}
{Meurer} G.~R.,  {Carignan} C.,  {Beaulieu} S.~F.,   {Freeman} K.~C.,  1996, \mn@doi [\aj] {10.1086/117895}, \href {https://ui.adsabs.harvard.edu/abs/1996AJ....111.1551M} {111, 1551}

\bibitem[\protect\citeauthoryear{{Narayanan}, {Anagha A.}  \& {Banerjee}}{{Narayanan} et~al.}{2024}]{ganesh_2024}
{Narayanan} G.,  {Anagha A.} G.,   {Banerjee} A.,  2024, \mn@doi [arXiv e-prints] {10.48550/arXiv.2407.02916}, \href {https://ui.adsabs.harvard.edu/abs/2024arXiv240702916N} {p. arXiv:2407.02916}

\bibitem[\protect\citeauthoryear{{Navarro}, {Frenk}  \& {White}}{{Navarro} et~al.}{1996}]{NFW}
{Navarro} J.~F.,  {Frenk} C.~S.,   {White} S. D.~M.,  1996, \mn@doi [\apj] {10.1086/177173}, \href {https://ui.adsabs.harvard.edu/abs/1996ApJ...462..563N} {462, 563}

\bibitem[\protect\citeauthoryear{{Oh} et~al.,}{{Oh} et~al.}{2015}]{ooh_2015}
{Oh} S.-H.,  et~al., 2015, \mn@doi [\aj] {10.1088/0004-6256/149/6/180}, 149, 180

\bibitem[\protect\citeauthoryear{{Ohta}, {Hamabe}  \& {Wakamatsu}}{{Ohta} et~al.}{1990}]{(OhtaHamabeandWakamatsu1990}
{Ohta} K.,  {Hamabe} M.,   {Wakamatsu} K.-I.,  1990, \mn@doi [\apj] {10.1086/168892}, \href {https://ui.adsabs.harvard.edu/abs/1990ApJ...357...71O} {357, 71}

\bibitem[\protect\citeauthoryear{Patra \& Jog}{Patra \& Jog}{2019}]{PatraJog2019}
Patra N.~N.,  Jog C.~J.,  2019, Monthly Notices of the Royal Astronomical Society

\bibitem[\protect\citeauthoryear{{Pe{\~n}arrubia}, {Navarro}, {McConnachie}  \& {Martin}}{{Pe{\~n}arrubia} et~al.}{2009}]{Peñarrubiaetal2009}
{Pe{\~n}arrubia} J.,  {Navarro} J.~F.,  {McConnachie} A.~W.,   {Martin} N.~F.,  2009, \mn@doi [\apj] {10.1088/0004-637X/698/1/222}, \href {https://ui.adsabs.harvard.edu/abs/2009ApJ...698..222P} {698, 222}

\bibitem[\protect\citeauthoryear{{Peebles}}{{Peebles}}{1969}]{peebles1969}
{Peebles} P.~J.~E.,  1969, \mn@doi [\apj] {10.1086/149876}, \href {https://ui.adsabs.harvard.edu/abs/1969ApJ...155..393P} {155, 393}

\bibitem[\protect\citeauthoryear{{Perret}}{{Perret}}{2016}]{Perret2016}
{Perret} V.,  2016, {DICE: Disk Initial Conditions Environment}, Astrophysics Source Code Library, record ascl:1607.002 (\mn@eprint {ascl} {1607.002})

\bibitem[\protect\citeauthoryear{{Peters} \& {Kuzio de Naray}}{{Peters} \& {Kuzio de Naray}}{2019}]{PetersandKuziodeNaray2019}
{Peters} W.,  {Kuzio de Naray} R.,  2019, \mn@doi [\mnras] {10.1093/mnras/sty3505}, \href {https://ui.adsabs.harvard.edu/abs/2019MNRAS.484..850P} {484, 850}

\bibitem[\protect\citeauthoryear{{Pfenniger}, {Saha}  \& {Wu}}{{Pfenniger} et~al.}{2023}]{Daniel2023}
{Pfenniger} D.,  {Saha} K.,   {Wu} Y.-T.,  2023, \mn@doi [\aap] {10.1051/0004-6361/202245463}, \href {https://ui.adsabs.harvard.edu/abs/2023A&A...673A..36P} {673, A36}

\bibitem[\protect\citeauthoryear{Pontzen \& Governato}{Pontzen \& Governato}{2013}]{Pontzen2013}
Pontzen A.,  Governato F.,  2013, \mn@doi [Monthly Notices of the Royal Astronomical Society] {10.1093/mnras/sts529}, 430, 121

\bibitem[\protect\citeauthoryear{Prada, Forero-Romero, Grand, Pakmor  \& Springel}{Prada et~al.}{2019}]{Prada_2019}
Prada J.,  Forero-Romero J.~E.,  Grand R. J.~J.,  Pakmor R.,   Springel V.,  2019, \mn@doi [Monthly Notices of the Royal Astronomical Society] {10.1093/mnras/stz2873}, 490, 4877

\bibitem[\protect\citeauthoryear{{Roberts}, {Huntley}  \& {van Albada}}{{Roberts} et~al.}{1979}]{Roberts1979}
{Roberts} W.~W. J.,  {Huntley} J.~M.,   {van Albada} G.~D.,  1979, \mn@doi [\apj] {10.1086/157367}, \href {https://ui.adsabs.harvard.edu/abs/1979ApJ...233...67R} {233, 67}

\bibitem[\protect\citeauthoryear{Rosas-Guevara et~al.,}{Rosas-Guevara et~al.}{2022}]{Rosas-Guevara2021}
Rosas-Guevara Y.,  et~al., 2022, Monthly Notices of the Royal Astronomical Society, 512, 5339

\bibitem[\protect\citeauthoryear{Saha \& Naab}{Saha \& Naab}{2013}]{Saha_2013}
Saha K.,  Naab T.,  2013, \mn@doi [Monthly Notices of the Royal Astronomical Society] {10.1093/mnras/stt1088}, 434, 1287–1299

\bibitem[\protect\citeauthoryear{{Saha} et~al.,}{{Saha} et~al.}{2010}]{Sahaetal2010}
{Saha} A.,  et~al., 2010, \mn@doi [\aj] {10.1088/0004-6256/140/6/1719}, \href {https://ui.adsabs.harvard.edu/abs/2010AJ....140.1719S} {140, 1719}

\bibitem[\protect\citeauthoryear{Sellwood}{Sellwood}{2008}]{Sellwood_2008}
Sellwood J.~A.,  2008, \mn@doi [The Astrophysical Journal] {10.1086/586882}, 679, 379

\bibitem[\protect\citeauthoryear{{Sellwood} \& {Wilkinson}}{{Sellwood} \& {Wilkinson}}{1993}]{SellwoodWilkinson1993}
{Sellwood} J.~A.,  {Wilkinson} A.,  1993, \mn@doi [Reports on Progress in Physics] {10.1088/0034-4885/56/2/001}, \href {https://ui.adsabs.harvard.edu/abs/1993RPPh...56..173S} {56, 173}

\bibitem[\protect\citeauthoryear{{Sellwood}, {Shen}  \& {Li}}{{Sellwood} et~al.}{2019}]{Sellwood2019}
{Sellwood} J.~A.,  {Shen} J.,   {Li} Z.,  2019, \mn@doi [\mnras] {10.1093/mnras/stz1145}, \href {https://ui.adsabs.harvard.edu/abs/2019MNRAS.486.4710S} {486, 4710}

\bibitem[\protect\citeauthoryear{Sheth, Regan, Scoville  \& Strubbe}{Sheth et~al.}{2003}]{Sheth_2003}
Sheth K.,  Regan M.~W.,  Scoville N.~Z.,   Strubbe L.~E.,  2003, \mn@doi [The Astrophysical Journal] {10.1086/377329}, 592, L13

\bibitem[\protect\citeauthoryear{{Shu}}{{Shu}}{1992}]{ShuFrankH1992}
{Shu} F.~H.,  1992, \mnras

\bibitem[\protect\citeauthoryear{{Silva-Lima}, {Martins}, {Coelho}  \& {Gadotti}}{{Silva-Lima} et~al.}{2022}]{SilvaLima2022}
{Silva-Lima} L.~A.,  {Martins} L.~P.,  {Coelho} P. R.~T.,   {Gadotti} D.~A.,  2022, \mn@doi [\aap] {10.1051/0004-6361/202142432}, \href {https://ui.adsabs.harvard.edu/abs/2022A&A...661A.105S} {661, A105}

\bibitem[\protect\citeauthoryear{{Sorensen}, {Matsuda}  \& {Fujimoto}}{{Sorensen} et~al.}{1976}]{Sorensen1976}
{Sorensen} S.~A.,  {Matsuda} T.,   {Fujimoto} M.,  1976, \mn@doi [\apss] {10.1007/BF00640025}, \href {https://ui.adsabs.harvard.edu/abs/1976Ap&SS..43..491S} {43, 491}

\bibitem[\protect\citeauthoryear{Sormani, Binney  \& Magorrian}{Sormani et~al.}{2015}]{stv1135}
Sormani M.~C.,  Binney J.,   Magorrian J.,  2015, \mn@doi [Monthly Notices of the Royal Astronomical Society] {10.1093/mnras/stv1135}, 451, 3437

\bibitem[\protect\citeauthoryear{Storchi-Bergmann \& Schnorr-Müller}{Storchi-Bergmann \& Schnorr-Müller}{2019}]{Storchi_Bergmann_2019}
Storchi-Bergmann T.,  Schnorr-Müller A.,  2019, \mn@doi [Nature Astronomy] {10.1038/s41550-018-0611-0}, 3, 48–61

\bibitem[\protect\citeauthoryear{Sørensen, Matsuda  \& Fujimoto}{Sørensen et~al.}{1976}]{Sørensen1976}
Sørensen S.,  Matsuda T.,   Fujimoto M.,  1976, \mn@doi [Astrophysics and Space Science] {10.1007/BF00640025}, 43, 491

\bibitem[\protect\citeauthoryear{{Teyssier}}{{Teyssier}}{2002}]{Tessier2002}
{Teyssier} R.,  2002, \aap, 385, 337

\bibitem[\protect\citeauthoryear{{Toomre}}{{Toomre}}{1964}]{Toomre}
{Toomre} A.,  1964, \mn@doi [\apj] {10.1086/147861}, \href {https://ui.adsabs.harvard.edu/abs/1964ApJ...139.1217T} {139, 1217}

\bibitem[\protect\citeauthoryear{Wadsley, Stadel  \& Quinn}{Wadsley et~al.}{2004}]{WADSLEY2004137}
Wadsley J.,  Stadel J.,   Quinn T.,  2004, \mn@doi [New Astronomy] {https://doi.org/10.1016/j.newast.2003.08.004}, 9, 137

\bibitem[\protect\citeauthoryear{{Weinberg}}{{Weinberg}}{1985a}]{Weinberg1985}
{Weinberg} M.~D.,  1985a, \mn@doi [\mnras] {10.1093/mnras/213.3.451}, \href {https://ui.adsabs.harvard.edu/abs/1985MNRAS.213..451W} {213, 451}

\bibitem[\protect\citeauthoryear{Weinberg}{Weinberg}{1985b}]{Weinberg_1985}
Weinberg M.~D.,  1985b, \mn@doi [Monthly Notices of the Royal Astronomical Society] {10.1093/mnras/213.3.451}, 213, 451

\bibitem[\protect\citeauthoryear{{Wozniak} \& {Pierce}}{{Wozniak} \& {Pierce}}{1991}]{WP}
{Wozniak} H.,  {Pierce} M.~J.,  1991, \aaps, \href {https://ui.adsabs.harvard.edu/abs/1991A&AS...88..325W} {88, 325}

\bibitem[\protect\citeauthoryear{{Wozniak}, {Friedli}, {Martinet}, {Martin}  \& {Bratschi}}{{Wozniak} et~al.}{1995}]{Wozniaketal1995}
{Wozniak} H.,  {Friedli} D.,  {Martinet} L.,  {Martin} P.,   {Bratschi} P.,  1995, \aaps, \href {https://ui.adsabs.harvard.edu/abs/1995A&AS..111..115W} {111, 115}

\bibitem[\protect\citeauthoryear{{de Vaucouleurs}}{{de Vaucouleurs}}{1959}]{deVaucouleurs1959}
{de Vaucouleurs} G.,  1959, \mn@doi [Handbuch der Physik] {10.1007/978-3-642-45932-0_8}, \href {https://ui.adsabs.harvard.edu/abs/1959HDP....53..311D} {53, 311}

\makeatother
\end{thebibliography}


\end{document}